\documentclass[%
 reprint,
 amsmath,amssymb,
 aps,
onecolumn]{revtex4-2} 

\usepackage{graphicx}
\usepackage{dcolumn}
\usepackage{bm}
\usepackage{latexsym}
\usepackage{amsmath,amssymb,enumerate,color,url}
\usepackage{theorem}
\usepackage{url}
\usepackage{algorithmic}
\usepackage{array}
\usepackage[tight,footnotesize]{subfigure}
\usepackage[font=footnotesize]{subfig}
\usepackage{color}
\usepackage{comment}
\newcommand{\dd}{\mathrm{d}}

\newcommand{\ii}{\mathrm{i}}

\theorembodyfont{\rmfamily}

\def\Hat#1{\widehat#1}

\begin{document}

\title{Derivation and Characteristics of Closed-Form Solutions of the Fundamental Equations for Online User Dynamics
}


\author{T.~Ikeya}%
\affiliation{%
Graduate School of Systems Design, Tokyo Metropolitan University, Japan. 
}%
\author{Masaki~Aida}%
\email[]{aida@tmu.ac.jp}
\affiliation{%
Graduate School of Systems Design, Tokyo Metropolitan University, Japan. 
}%



\begin{abstract}
The oscillation model, based on the wave equation on networks, can describe user dynamics in online social networks.
The fundamental equation of user dynamics can be introduced into the oscillation model to explicitly describe the causal relation of user dynamics yielded by certain specific network structures.
Moreover, by considering the sparseness of the link structure of online social networks, a novel fundamental equation of different forms has been devised.
In this paper, we derive a closed-form solution of the new fundamental equation.
Also, we show that the closed-form solution of the new fundamental equation can generate the general solution of the original wave equation and investigate the characteristics of the derived general solution. 
\end{abstract}

\pacs{64.60.aq}
\keywords{online social networks, oscillation model, user dynamics, wave equation}

\maketitle

\section{Introduction}
\label{sec.1}
In recent years, people around the world have been increasingly using social networking services (SNSs).
With the widespread use of SNSs, people can easily transmit and exchange information in online social networks (OSNs).
Accordingly, SNSs can bring benefits to people in the virtual space to facilitate the enrichment of friendships and collect information about their preferences.

However, the SNSs also have their negative aspects.
Prime examples are the online flaming phenomenon caused by collective user dynamics and the online echo-chamber phenomenon in which specific opinions and beliefs are strengthened among people in a closed community.
These phenomena are likely to negatively impact not only on online communities such as SNSs but also on activities in the real world.
To consider countermeasures to such phenomena, it is necessary to understand the characteristics of user dynamics caused by interactions between users via OSNs.

As one of the models created to describe user dynamics in OSNs, the previous study \cite{oscillation_model,oscillation_model2} proposed an oscillation model based on the wave equation.
Additionally, in the framework of the oscillation model, the fundamental equations that can explicitly describe the causal relationship between user dynamics and certain specific network structures were introduced \cite{fundamental_equation_1,fundamental_equation_2}.
These fundamental equations of OSNs are of two types: one cannot consider the link structure whereas the other can.
The first type of fundamental equation is easy to solve and we can obtain closed-form solutions, but no closed-form solutions of the second type of fundamental equation have been published so far.
This paper is an extended vesiion of a conference paper~\cite{ikeya} and derives a closed-form solution of the second type of fundamental equation for OSNs.
We also show the closed-form solution of the second type of fundamental equation can generate the general solution of the original wave equation. 
In addition, we show the characteristics of the generated solution, and clarify it is important for describing causality of online user dynamics. 

The rest of this paper is organized as follows.
In Sec.~\ref{sec.2}, we describe studies on SNS and user dynamics related to this work and explain innovation in our research.
In Sec.~\ref{sec.3}, we overview the oscillation model based on the wave equation on networks and describe two different fundamental equations, they are called boson-type and fermion-type equations. 
The boson-type equation cannot consider the link structure of OSNs, but the fermion-type equation can. 
In Sec.~\ref{sec.4}, we derive a closed-form expression of the fermion-type fundamental equation. 
In Sec.~\ref{sec.5}, we show the closed-form solutions of the fermion-type fundamental equation can generate a general solution of the original wave equation.
In Sec.~\ref{sec.6}, we show the characteristics of a general solution of the original wave equation derived from the closed-form solutions of the fermion-type fundamental equation, and clarify it is important for describing the causality of online user dynamics.
In Sec.~\ref{sec.7}, we present our conclusions.

\section{Related Work}
\label{sec.2}
Explosive user dynamics in OSNs, including online flaming phenomena, can seriously impact not only online communities but also social activities in the real world. 
Therefore, understanding the online user dynamics is an important issue.
Such online user dynamics are generated from user interactions and manifest themselves as a divergence of the intensity of online user activity.

Studies on user dynamics in OSNs have examined by various models in an effort to capture the diversity of the characteristics of user dynamics. 
User dynamics that describe the adoption and abandonment of a particular SNS have been modeled by the SIR model, which is a traditional epidemiological model, and the irSIR model, which is an extension of that model~\cite{SIR,Nekovee2007,Cannarella2014EpidemiologicalMO}. 
These models express the state transition of an objective system in a macroscopic framework but are not good at describing individual user dynamics. 
Also, these models describe the speed of changes in transient states and/or the configuration of the final steady-state of the system, but divergence in the intensity of user dynamics is not addressed by these models. 

The consensus problem including user opinion formation is typical of the dynamics in OSNs~\cite{Olfati-Saber2004,Wang2010}. 
This can be modeled by a first-order differential equation with respect to time using a Laplacian matrix that represents the social network structure.
The differential equation used in this model is a sort of continuous-time Markov chain on the network.
First-order differential equations with respect to time are also used in modeling of the temporal change in social network structure (linking or delinking of the nodes), and there are models that consider change via a continuous-time Markov chain~\cite{Snijders2010}. 
In addition to theoretical models, \cite{Cha2009} and \cite{Zhao2012} studied user dynamics analysis based on real network observations. 
Similar to epidemiological models, the Markov chain describes the transient states and/or the steady-state of the system, but not the divergence of user dynamics. 

\begin{figure}[tb]
  \centering
\includegraphics[width=0.5\linewidth]{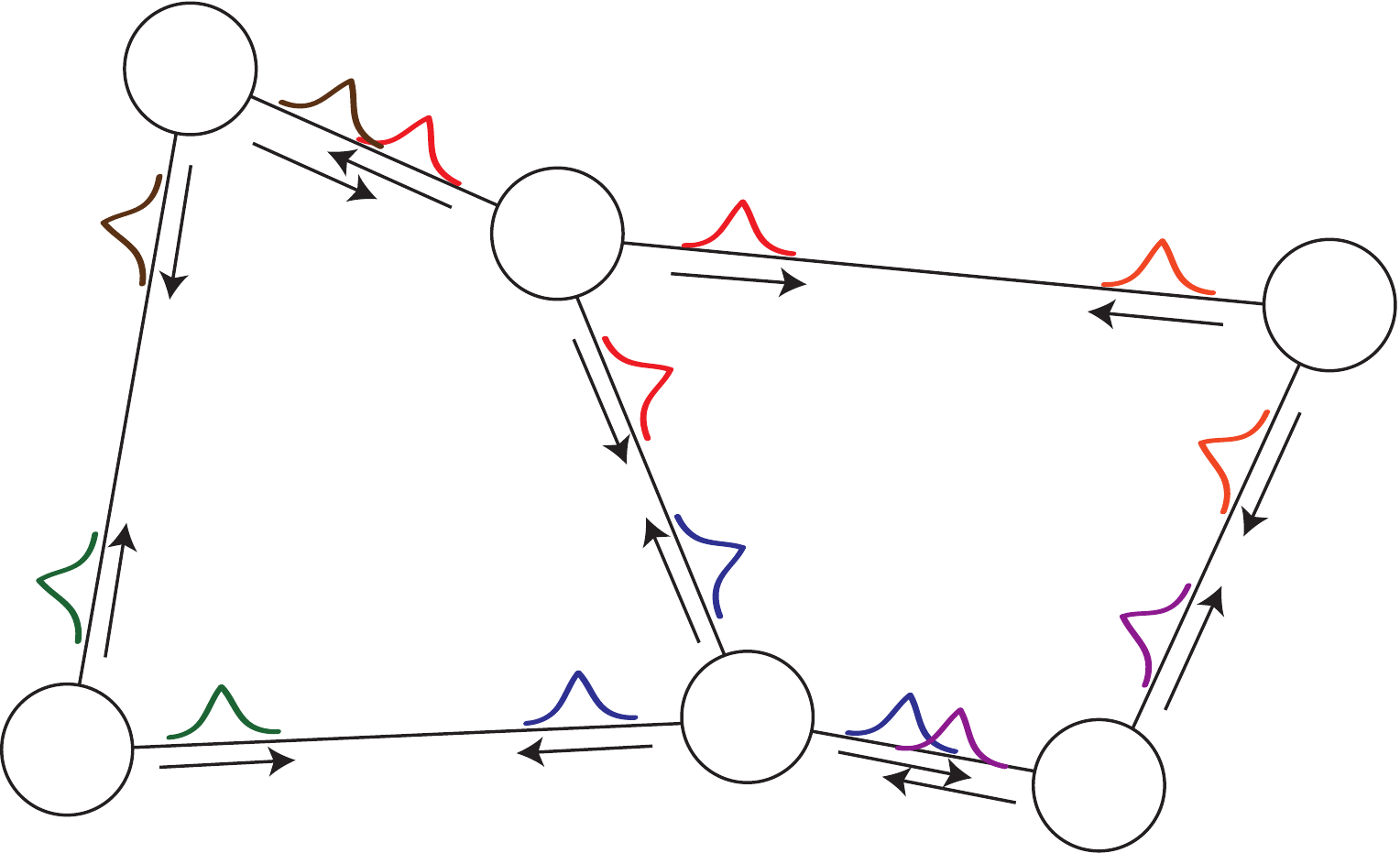}
    \caption{Propagation of influences between users through links of OSNs}
    \label{fig:propagation}
\end{figure}

The oscillation model, the main topic in this paper, has been introduced for describing explosive user dynamics in OSNs including online flaming phenomena \cite{oscillation_model}.
It is based on the wave equation on networks. 
Although the wave equation may not be familiar in the field of network engineering,  it can describe the propagation of a certain influence between users at a {\it finite speed}.
In this case, the wave equation describes the propagation of influence between users through links of OSNs. 
Figure~\ref{fig:propagation} illustrates such a situation, where nodes denote users and links denote the relationships between users. 

Let us consider a waveform of the influence between users. 
Since the influence propagates at a finite speed, the waveform propagates at a finite speed. 
Also, the waveform can be decomposed into sine waves by the Fourier transform because any continuous function can be. 
Therefore, the decomposed sine waves propagate also at a finite speed (see Fig.~\ref{fig:propagation}). 
This is the reason why the online user dynamics generated by interactions between users can be described by the wave equation-based model. 

\begin{figure}[bt]
  \centering
\includegraphics[width=0.4\linewidth]{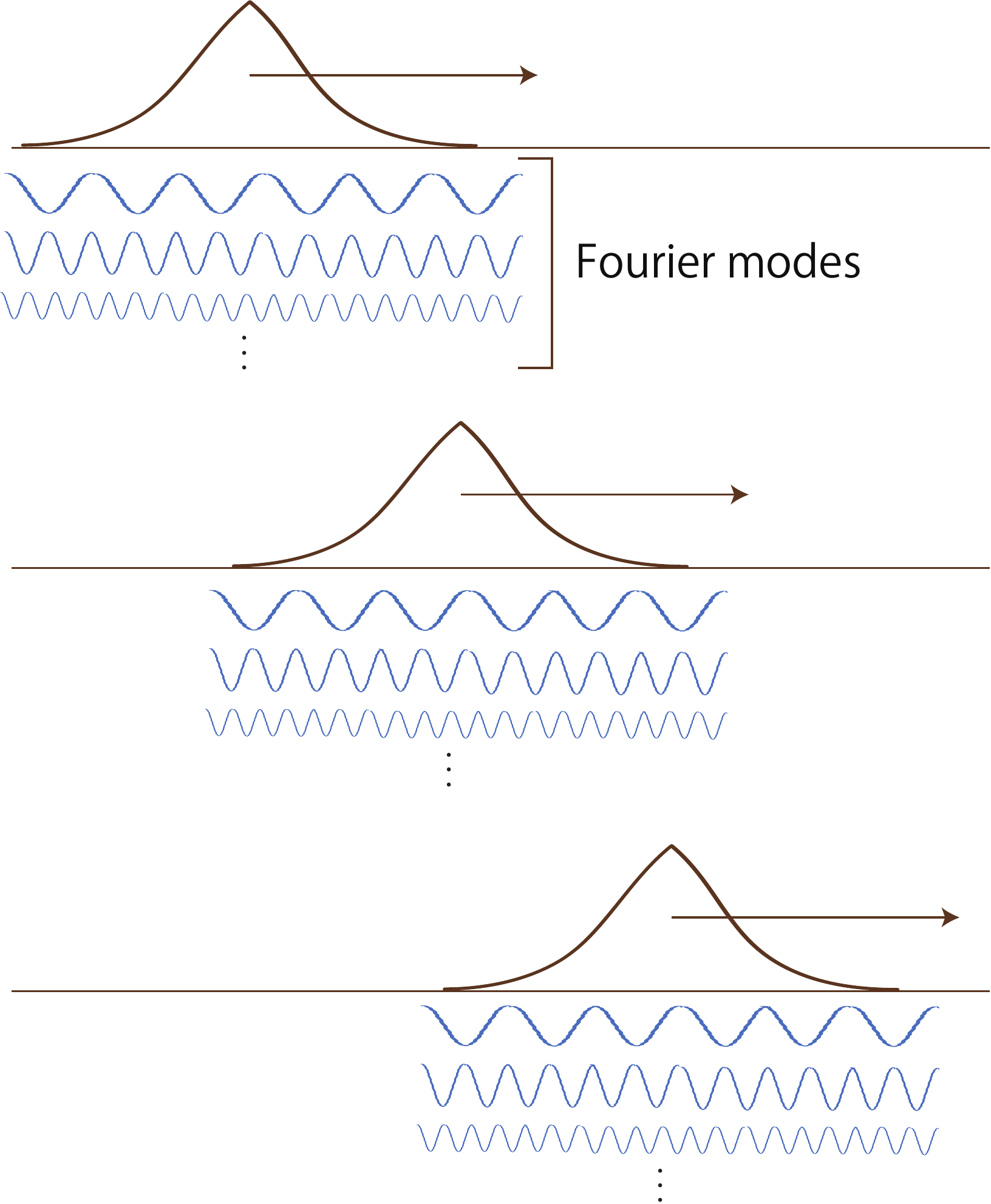}
    \caption{Wave equation in online social networks}
    \label{fig:wave-eq}
\end{figure}

The validity of the oscillation model is also supported by the fact that the oscillation energy of the node represents the well-known node centrality.
The oscillation energy calculated from the oscillation model gives the conventional node centrality (degree centrality and betweenness centrality) in simple cases and also gives a generalized notion of node centrality in various situations~\cite{takano,takano2}. 
In addition, the oscillation model can describe explosive user dynamics including online flaming phenomena as divergence in the oscillation energy. 

Moreover, the oscillation model yields fundamental equations that can describe not only user dynamics but also causal relationships between user dynamics and the structure of OSNs \cite{fundamental_equation_1,fundamental_equation_2}.

There are two types of fundamental equations: boson-type and fermion-type \cite{ITC2020}. 
The boson-type fundamental equation cannot consider the structure of links, whereas the fermion-type can take account of characteristics such as the sparseness of the link structure of OSNs.
The solutions of both types of fundamental equations can generate general solutions of the original wave equation of OSNs. 

For the boson-type fundamental equation, the closed-form solution is easily obtained, but the closed-form solution for the fermion-type equation is not obtained yet.
Since the fermion-type fundamental equation is more suitable for describing user dynamics in actual OSN structures, derivation of closed-form solutions of the fermion-type fundamental equation remains the desired target. 

In this paper, we derive a closed-form solution of the fermion-type fundamental equation.
Also, we show general solutions of the original wave equation generated from the closed-form solution of the fermion-type fundamental equation.
In addition, we show the characteristics of a general solution of the original wave equation derived from the closed-form solutions of the fermion-type fundamental equation and clarify it is important for describing the causality of online user dynamics.

\section{Oscillation Model for Online Social Networks}
\label{sec.3}
This section briefly explains the oscillation model \cite{oscillation_model} for describing user dynamics in OSNs, and introduces two types of fundamental equations \cite{fundamental_equation_1}.  

\subsection{Wave Equation-Based Model for Online User Dynamics}
Let $G(V,\, E)$ be a directed graph representing the structure of an OSN with $n$ nodes ($n$ users), where 
$V=\{ 1, \, 2, \, \dots, \, n \}$ denotes the set of nodes and $E$ denotes the set of links (the relationship between users).
For a directed link from node $i$ to node $j$, $(i \rightarrow j) \in E$, the weight of directed link $(i \rightarrow j)$ is denoted by $w_{ij}>0$; the adjacency matrix $\bm{\mathcal{A}}:=[\mathcal{A}_{i \, j}]_{1 \leq i, \, j \leq n}$ is defined as
\begin{align*}
  \mathcal{A}_{ij} :=
  \begin{cases}
    w_{ij}, & ((i\rightarrow j) \in E),\\
    0, & ((i\rightarrow j) \notin E).
  \end{cases}
\end{align*}
Furthermore, the weighted nodal degree of node $i$ is defined as $d_{i}:=\sum_{j \in \partial i}w_{ij}$ and the degree matrix is given as 
\[
\bm{\mathcal{D}}:={\rm diag}(d_{1}, \, d_{2}, \, \dots, \, d_{n}),
\]
where $\partial i$ is the set of nodes adjacent from node $i$.
Also, the Laplacian matrix of $\mathcal{G}(V,\, E)$ is defined as 
\[
\bm{\mathcal{L}}:=\bm{\mathcal{D}}-\bm{\mathcal{A}}.
\]

The oscillation model is a minimal model for describing user dynamics, that is, we assume a universal model as simple as possible.
First, assuming that the state of each node can be described by a simple one-dimensional variable, we let $x_{i}(t)$ be the state of node $i$ at time $t$ and $\bm{x}(t):={}^t\!(x_{1}(t), \, x_{2}(t), \, \dots, \, x_{n}(t))$ be an $n$-dimensional state vector for all nodes.
Next, we introduce the interaction between nodes. 
Between adjacent nodes $i$ and $j$, a force acts in a direction so as to reduce the difference in node state between nodes $i$ and $j$.
The strength of this force is proportional to the absolute value of the difference in state quantities: $|x_{i}(t) - x_{j}(t)|$.
The equation of motion of the user state vector in OSNs is expressed as follows: 
\begin{align}
  \frac{{\rm d}^2}{{\rm d}t^{2}}\, \bm{x}(t) = -\bm{\mathcal{L}} \, \bm{x}(t).
  \label{eom}
\end{align}
This is called the wave equation on networks.
The wave equation (\ref{eom}) describes a phenomenon that inter-user influence propagates through OSNs at a finite speed.

\subsection{Fundamental Equations of Online User Dynamics}
The fundamental equations have been introduced for describing the causal relationship between user dynamics and the structure of OSNs. 
This means that we can understand the causal relationship of a certain specific structure of networks on user dynamics. 
Here, understanding the causal relationship is the following situation. 
Consider the situation that we completely know the relationship between the structure of an OSN and user dynamics generated from the OSN. 
If we change the structure of the OSN by adding some links to the OSN, the difference of the newly generated user dynamics can be understood by using a newly added graph structure. 
To describe the causal relationship, it is necessary to describe user dynamics by a first-order differential equation with respect to time \cite{oscillation_model,fundamental_equation_1}. 
Since the wave equation (\ref{eom}) is a second-order differential equation with respect to time, it cannot describe the causal relations. 
The fundamental equations shown later are first-order differential equations with respect to time and can describe user dynamics and causal relations.

Let us consider OSNs whose Laplacian matrix has only real eigenvalues and is diagonalizable. 
This constraint is needed because non-real eigenvalues of the Laplacian matrix cause the divergence of oscillation energy like online flaming phenomena. 
We now concentrate on the situations that the strength of user dynamics does not diverge. 
In this case, the Laplacian matrix is a semi-positive definite matrix, and its square root is uniquely determined as a semi-positive definite matrix. 
Let the square root matrix of $\bm{\mathcal{L}}$ be $\sqrt{\bm{\mathcal{L}}}$, which is an $n\times n$ matrix and $\big(\sqrt{\bm{\mathcal{L}}}\big)^2 = \bm{\mathcal{L}}$. 
Using $\sqrt{\bm{\mathcal{L}}}$, the fundamental equations are expressed as 
\begin{align}
  \pm \ii \, \frac{\dd}{\dd t} \, \bm{x}^{\pm}(t)
  =\sqrt{\bm{\mathcal{L}}} \, \bm{x}^{\pm}(t),
  \  \text{(double sign corresponds)},
  \label{fundamental_equation_eom}
\end{align}
where $\bm{x}^{\pm}(t)$ is an $n$-dimensional vector. 
The solutions $\bm{x}^{\pm}(t)$ of \eqref{fundamental_equation_eom} are also solutions of the original wave equation (\ref{eom}), because 
\[
\frac{\dd^2}{\dd t^2} \, \bm{x}^{\pm}(t) = \mp\ii\,\frac{\dd}{\dd t} \, \sqrt{\bm{\mathcal{L}}} \, \bm{x}^{\pm}(t) = -\bm{\mathcal{L}} \, \bm{x}^{\pm}(t).
\]

\begin{figure}[bt]
  \centering
    \includegraphics[width=0.7\linewidth]{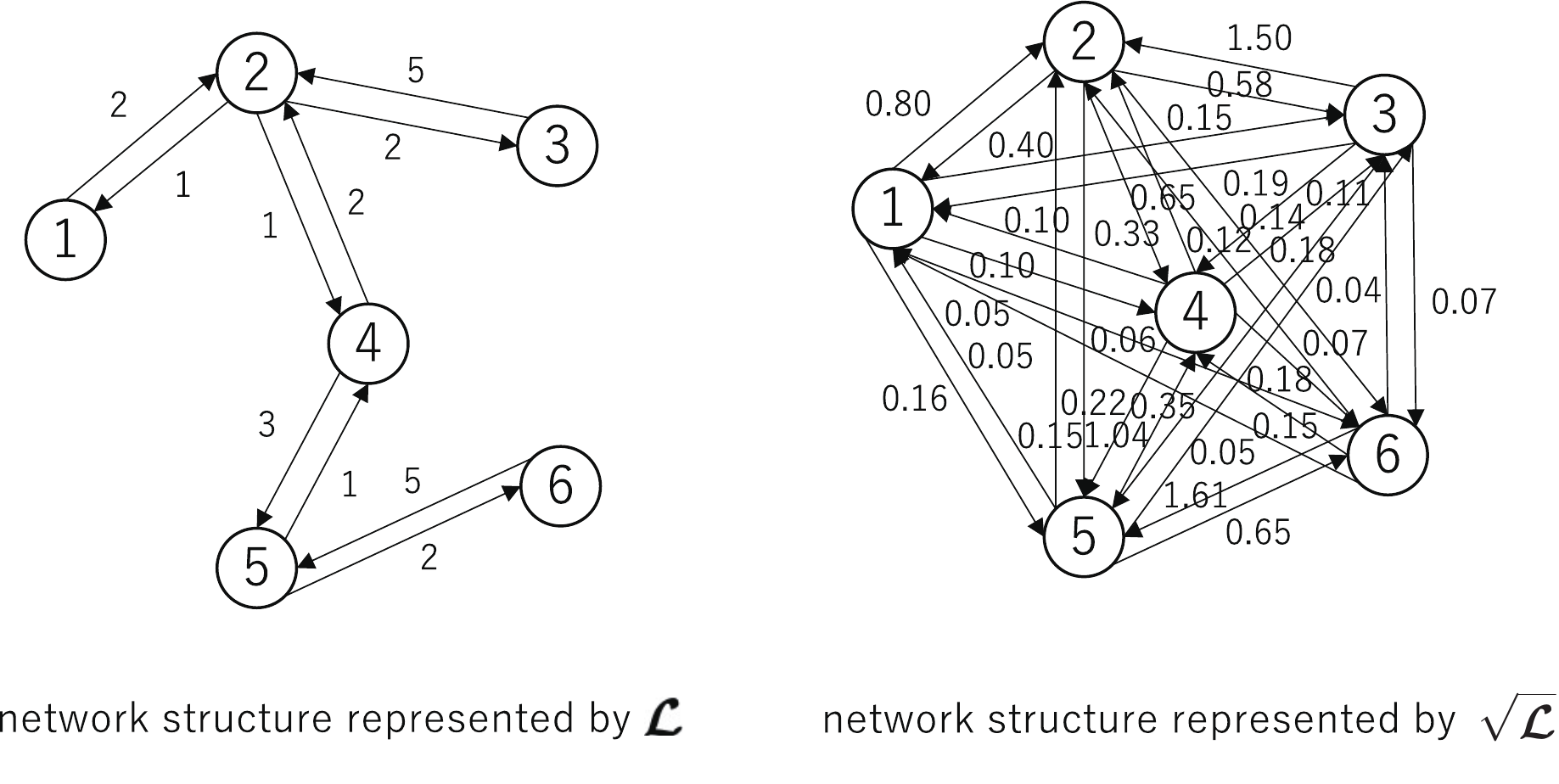}
    \caption{Network structures represented by $\bm{\mathcal{L}}$ and $\sqrt{\bm{\mathcal{L}}}$}
    \label{laplacian_and_root}
\end{figure}

By introducing $2n$-dimensional vector 
\[
\bm{\Hat{x}}(t) := \bm{x}^{+}(t)\otimes
  \begin{pmatrix}
    1\\
    0
  \end{pmatrix}+\bm{x}^{-}(t)\otimes
  \begin{pmatrix}
    0\\
    1
  \end{pmatrix},
\]
the fundamental equation (\ref{fundamental_equation_eom}) is expressed by the following one equation: 
\begin{align}
  {\rm i} \, \frac{{\rm d}}{{\rm d}t}  \, \bm{\hat{x}}(t) = \left(\sqrt{\bm{\mathcal{L}}} \otimes 
  \begin{bmatrix}
  +1 & 0\\
  0 & -1
\end{bmatrix}
\right) \bm{\hat{x}}(t),
 \label{fundamental_equation_eom_two}
\end{align}
where $\otimes$ denotes the Kronecker product \cite{kronecker}.
We call the fundamental equation \eqref{fundamental_equation_eom_two} (and equivalently \eqref{fundamental_equation_eom}) the boson-type fundamental equation. 

The closed-form solution of the fundamental equation~(\ref{fundamental_equation_eom}) is easily obtained as 
\begin{align}
\bm{x}^\pm(t) &= \bm{P} \, \exp(\mp\ii\,\bm{\Omega}\,t) \, \bm{P}^{-1} \, \bm{x}^\pm(0), 
\quad \text{(double-sign corresponds)}. 
\label{solution-x-pm}
\end{align}
where $\bm{P}$ is a regular matrix consisting of the eigenvectors of the Laplacian matrix $\bm{\mathcal{L}}$. 
Also, $\bm{\Omega}$ is a diagonal matrix, and is defined as follows. 
First, we define  the following diagonal matrix 
\[
\bm{\Lambda} := \text{diag}(\lambda_0,\,\lambda_1,\,\dots,\,\lambda_{n-1}),
\]
where$\lambda_\mu$ $(\mu=0,\,1,\,\dots,\,n-1)$ denotes eigenvalues of $\bm{\mathcal{L}}$ and $\lambda_0 = 0$ (the Laplacian matrix has at keast one $0$ eigenvalue).
$\bm{\mathcal{L}}$ is diagonalized as  
\[
\bm{\Lambda} =  \bm{P}^{-1} \,  \bm{\mathcal{L}} \, \bm{P}. 
\]
In addition, 
\[
\bm{\Omega} := \sqrt{\bm{\Lambda}} = \mathrm{diag}(\omega_0,\,\omega_1,\,\dots,\,\omega_{n-1}),
\]
and, for eigenvalue $\lambda_\mu$ of $\bm{\mathcal{L}}$, $\omega_\mu = \sqrt{\lambda_\mu}$ and $\omega_0=0$. 
Rewriting the above solutions as the solution of the fundamental equation~(\ref{fundamental_equation_eom_two}) is expressed as follows:
\begin{align}
\bm{\hat{x}}(t) &= \left(\bm{P} \, \exp(-\ii\,\bm{\Omega}\,t) \, \bm{P}^{-1} \otimes 
\begin{bmatrix}
+1 & 0\\
0 & 0
\end{bmatrix}
+ \bm{P} \, \exp(+\ii\,\bm{\Omega}\,t) \, \bm{P}^{-1} \otimes 
\begin{bmatrix}
0 & 0\\
0 & +1
\end{bmatrix}
\right)\, \bm{\Hat{x}}(0) 
\label{solution-sqrt{L}}.
\end{align}

Although the boson-type fundamental equation \eqref{fundamental_equation_eom_two} can describe both user dynamics and the causal relation, they have an unacceptable problem.
The link structure of $\sqrt{\bm{\mathcal{L}}}$ is generally a complete graph even if the network represented by $\bm{\mathcal{L}}$ has a sparse structure.
In OSNs, the situation that all users of the world are connected directly is obviously an unacceptable situation. 
Figure~\ref{laplacian_and_root} shows an example of such a situation. 
The left panel shows an example of OSNs with 6 nodes. 
The square root of the Laplacian matrix shown in the left panel corresponds to a complete graph shown in the right panel. 
This is an unrealistic situation because there are extra links set between unrelated users.
The matrix appearing in the fundamental equation should have exactly the same link structure as the original OSNs 

\begin{figure}[bt]
  \centering
    \includegraphics[width=0.7\linewidth]{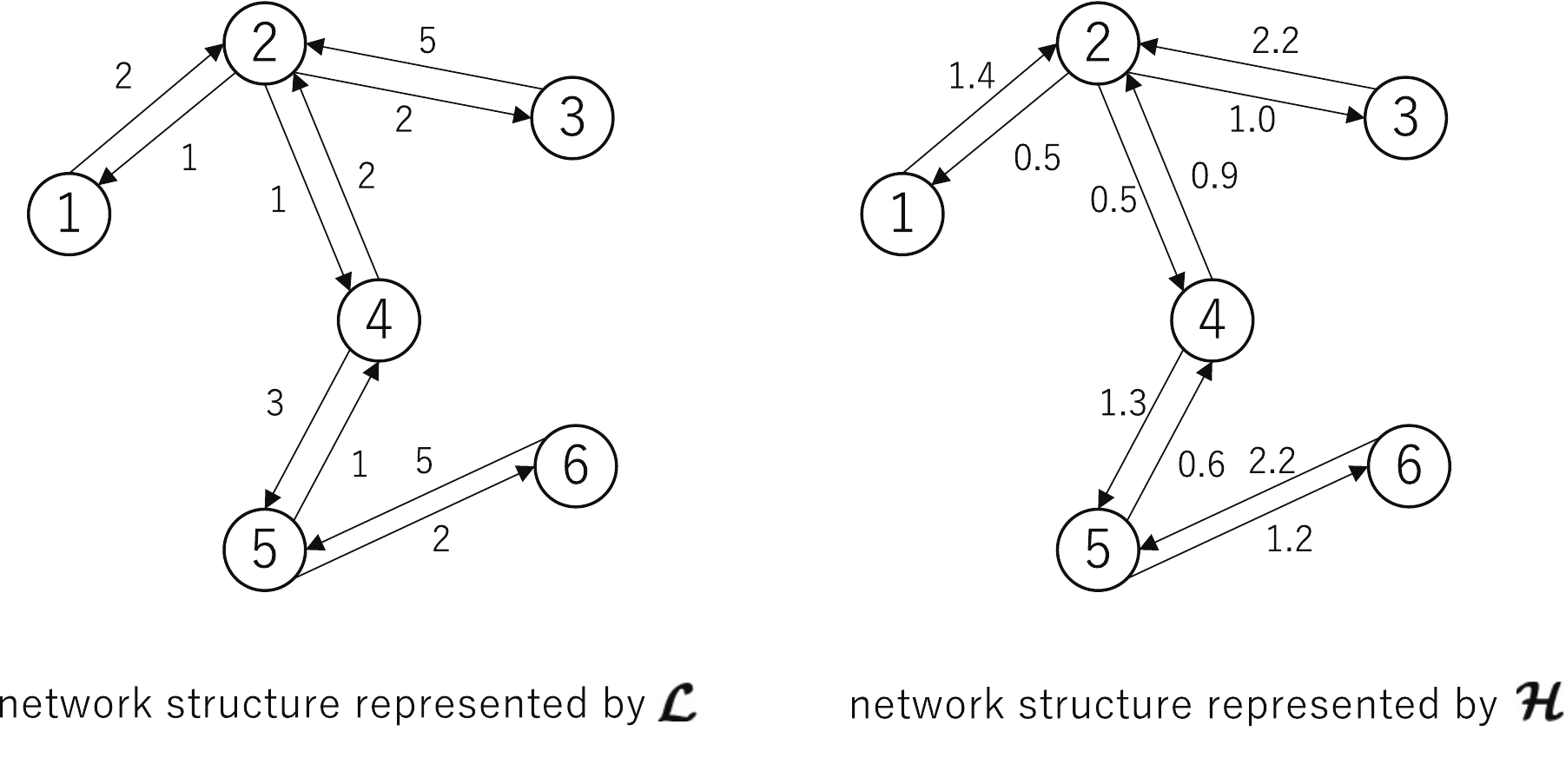}
    \caption{Network structures represented by $\bm{\mathcal{L}}$ and $\bm{\mathcal{H}}$}
    \label{laplacian_and_root_2}
\end{figure}

The fermion-type fundamental equation can avoid the above problem. 
First, we introduce the semi-normalized Laplacian matrix. 
As the well-known normalized Laplacian matrix is defined as
\[
\bm{\mathcal{N}} 
:= \sqrt{\bm{\mathcal{D}}^{-1}} \, \bm{\mathcal{L}} \, \sqrt{\bm{\mathcal{D}}^{-1}}
= \bm{I} - \sqrt{\bm{\mathcal{D}}^{-1}} \, \bm{\mathcal{A}} \, \sqrt{\bm{\mathcal{D}}^{-1}} ,
\]
let us define the semi-normalized Laplacian matrix as
\[
\bm{\mathcal{H}} 
:= \sqrt{\bm{\mathcal{D}}^{-1}} \, \bm{\mathcal{L}}
= \sqrt{\bm{\mathcal{D}}} - \sqrt{\bm{\mathcal{D}}^{-1}} \, \bm{\mathcal{A}},
\]
where $\bm{I}$ is the $n\times n$ unit matrix. 
Note that the semi-normalized Laplacian matrix is a kind of Laplacian matrix whose link weight of $(i\rightarrow j)$ is $w_{ij}/d_i$. 
In addition, the link structure of $\bm{\mathcal{H}}$ is completely the same as that of $\bm{\mathcal{L}}$ with respect to the presence or absence of links (see Fig.~\ref{laplacian_and_root_2}).

Second, let us introduce the $2n\times 2n$ matrix $\bm{\mathcal{\Hat{H}}}$ defined as 
\begin{align}
  \bm{\mathcal{\hat{H}}}:=\sqrt{\bm{\mathcal{D}}}\otimes
  \begin{bmatrix}
    +1 & 0\\
    0 & -1
  \end{bmatrix}
  -\left(\sqrt{\bm{\mathcal{D}}^{-1}} \, \bm{\mathcal{A}}\right)\otimes \frac{1}{2}
  \begin{bmatrix}
    +1 & +1\\
    -1 & -1
  \end{bmatrix},
  \label{h_hat_g}
\end{align}
where the $2\times 2$ matrix, the second term on the right side of \eqref{h_hat_g}, is a nilpotent matrix that becomes a zero matrix when squared.
We call $\bm{\mathcal{\Hat{H}}}$ the Hamiltonian in this paper. 
By using the Hamiltonian, the fermion-type fundamental equation is defined as 
\begin{align}
  {\rm i} \, \frac{{\rm d}}{{\rm d}t}  \, \bm{\Hat{x}}(t) = \bm{\mathcal{\Hat{H}}} \, \bm{\Hat{x}}(t). 
 \label{fundamental_equation_eom_3}
\end{align}

Since the fermion-type fundamental equation (\ref{fundamental_equation_eom_3}) is a first-order differential equation with respect to time, it can describe causal relationships. 
Also, with regard to user dynamics, the solutions of the fermion-type fundamental equation (\ref{fundamental_equation_eom_3}) can generate solutions of the original wave equation (\ref{eom}). 
Details are as follows. 
From the fermion-type fundamental equation (\ref{fundamental_equation_eom_3}), we obtain the second derivative of $\bm{\Hat{x}}(t)$ with respect to time $t$ as follows:
\begin{align}
  \frac{{\rm d}^{2}\bm{\Hat{x}}(t)}{{\rm d}t^{2}}
  &=-\bm{\mathcal{\hat{H}}}^{2} \, \bm{\hat{x}}(t)\notag\\
  &= -  \left(  \left[\bm{\mathcal{D}}
  - \frac{1}{2}\left(\bm{\mathcal{A}}+\sqrt{\bm{\mathcal{D}}^{-1}} \, \bm{\mathcal{A}} \, \sqrt{\bm{\mathcal{D}}}\right)\right]  \otimes \begin{bmatrix}
    +1 & 0\\
    0 & +1
  \end{bmatrix}
  -\frac{1}{2}\left(\bm{\mathcal{A}}-\sqrt{\bm{\mathcal{D}}^{-1}} \, \bm{\mathcal{A}} \, \sqrt{\bm{\mathcal{D}}}\right)\otimes
  \begin{bmatrix}
    0 & +1\\
    +1 & 0
  \end{bmatrix}\right) \, \bm{\Hat{x}}(t).
  \label{differential_two}
\end{align}
By extracting the differential equation for each of $\bm{x}^{+}(t)$ and $\bm{x}^{-}(t)$ from \eqref{differential_two}, we obtain
\begin{align}
  \frac{{\rm d}^{2}\bm{x}^{+}(t)}{{\rm d}t^{2}}
  &=- \left[\bm{\mathcal{D}}
  - \frac{1}{2}\left(\bm{\mathcal{A}}+\sqrt{\bm{\mathcal{D}}^{-1}} \, \bm{\mathcal{A}} \, \sqrt{\bm{\mathcal{D}}}\right)\right] \bm{x}^{+}(t)
  +\frac{1}{2}\left(\bm{\mathcal{A}}-\sqrt{\bm{\mathcal{D}}^{-1}} \, \bm{\mathcal{A}} \, \sqrt{\bm{\mathcal{D}}}\right) \, \bm{x}^{-}(t),
  \label{3.30}\\
  \frac{{\rm d}^{2}\bm{x}^{-}(t)}{{\rm d}t^{2}}
  &=- \left[\bm{\mathcal{D}}
  - \frac{1}{2}\left(\bm{\mathcal{A}}+\sqrt{\bm{\mathcal{D}}^{-1}} \, \bm{\mathcal{A}} \, \sqrt{\bm{\mathcal{D}}}\right)\right] \bm{x}^{-}(t)
  +\frac{1}{2}\left(\bm{\mathcal{A}}-\sqrt{\bm{\mathcal{D}}^{-1}} \, \bm{\mathcal{A}} \, \sqrt{\bm{\mathcal{D}}}\right) \, \bm{x}^{+}(t).
  \label{3.31}
\end{align}
Furthermore, by adding both sides of \eqref{3.30} and \eqref{3.31} respectively, we obtain
\begin{align}
  \frac{{\rm d}^{2}}{{\rm d}t^{2}} (\bm{x}^{+}(t) + \bm{x}^{-}(t)) &=
  - (\bm{\mathcal{D}} - \bm{\mathcal{A}}) \, (\bm{x}^{+}(t) + \bm{x}^{-}(t))
  \notag\\
  &=
  - \bm{\mathcal{L}} \, (\bm{x}^{+}(t) + \bm{x}^{-}(t)).
  \label{3.32}
\end{align}
This means that the sum of $\bm{x}^{+}(t)$ and $\bm{x}^{-}(t)$ are solutions of the original wave equation \eqref{eom}.

Finally, we give the Hamiltonian in the fermion-type fundamental equation (\ref{fundamental_equation_eom_3}) in a different form. 
The Hamiltonian (\ref{h_hat_g}) is equivalent to
\begin{align}
  \bm{\mathcal{\Hat{H}}}
  :=\bm{\mathcal{H}}\otimes \frac{1}{2}
  \begin{bmatrix}
    +1 & +1\\
    -1 & -1
  \end{bmatrix}
  +\sqrt{\bm{\mathcal{D}}}\otimes \frac{1}{2}
  \begin{bmatrix}
    +1 & -1\\
    +1 & -1
  \end{bmatrix}. 
  \label{Hamiltonian}
\end{align}
This is a convenient form for deriving solutions of the fundamental equation (\ref{fundamental_equation_eom_3}).

\section{Closed-Form Solution of the Fermion-Type Fundamental Equation}
\label{sec.4}
In this section, after preparing necessary algebraic relations, we derive a closed-form solution of the fermion-type fundamental equation (\ref{fundamental_equation_eom_3}).

\subsection{Preliminaries for the Algebraic Structure}
The solution of the fundamental equation (\ref{fundamental_equation_eom_3}) is formally expressed as 
\begin{align}
    \bm{\hat{x}}(t) &= \exp\!\left(-{\rm i} \,  \bm{\mathcal{\hat{H}}} \, t\right) \bm{\hat{x}}(0)
= \sum_{k=0}^\infty \frac{1}{k!} \left(-{\rm i} \,  \bm{\mathcal{\hat{H}}} \, t\right)^k \bm{\hat{x}}(0) .
  \label{solution_formsl}
\end{align}
Before deriving the closed-form solution, we introduce important algebraic relations. 
Let us define
\[
\bm{\hat{a}} := \frac{1}{2} \,
\begin{bmatrix}
+1 & +1\\
-1 & -1
\end{bmatrix}, \quad 
\bm{\hat{b}} := \frac{1}{2} \,
\begin{bmatrix}
+1 & -1\\
+1 & -1
\end{bmatrix}, \quad  
\bm{\hat{e}} :=
\begin{bmatrix}
+1 & 0\\
0 & +1
\end{bmatrix},
\]
where $\bm{\hat{a}}$ and $\bm{\hat{b}}$ appear in the Hamiltonian (\ref{Hamiltonian}) as 
\[
\bm{\mathcal{\Hat{H}}}
  :=\bm{\mathcal{H}}\otimes \bm{\hat{a}}
  +\sqrt{\bm{\mathcal{D}}}\otimes \bm{\hat{b}}. 
\]
They satisfy the following anticommutation relations: 
\begin{align}
\{\bm{\hat{a}},\bm{\hat{b}}\} := \bm{\hat{a}}\bm{\hat{b}} + \bm{\hat{b}}\bm{\hat{a}} = \bm{\hat{e}}, \quad \bm{\hat{a}}^2 = \bm{\hat{b}}^2 = \bm{\hat{\mathrm{o}}}\,\text{(null matrix)}. 
\end{align}
From the anticommutation relation of $\bm{\hat{a}}$ and $\bm{\hat{b}}$, we have 
\begin{align}
\bm{\hat{a}}\bm{\hat{b}}\bm{\hat{a}} &=(\bm{\hat{e}} - \bm{\hat{b}}\bm{\hat{a}})\, \bm{\hat{a}} =\bm{\hat{a}} - \bm{\hat{b}}\bm{\hat{a}}^2 = \bm{\hat{a}}, 
\notag\\
\bm{\hat{b}}\bm{\hat{a}}\bm{\hat{b}} &=(\bm{\hat{e}} - \bm{\hat{a}}\bm{\hat{b}}) \, \bm{\hat{b}} = \bm{\hat{b}} - \bm{\hat{a}} \bm{\hat{b}}^2 = \bm{\hat{b}}, 
\notag\\
\bm{\hat{a}}\bm{\hat{b}}\bm{\hat{a}}\bm{\hat{b}} &=(\bm{\hat{e}} -\bm{\hat{b}}\bm{\hat{a}}) \, \bm{\hat{a}}\bm{\hat{b}} = \bm{\hat{a}} \,\bm{\hat{b}}, 
\notag\\
\bm{\hat{b}}\bm{\hat{a}}\bm{\hat{b}}\bm{\hat{a}} &=(\bm{\hat{e}} - \bm{\hat{a}}\bm{\hat{b}}) \, \bm{\hat{b}}\bm{\hat{a}} = \bm{\hat{b}}\bm{\hat{a}} .
\notag
\end{align}
Therefore, when $\bm{\mathcal{\Hat{H}}}^k$ is expanded, the matrices that appear to the right of the Kronecker product are just 
\[
\bm{\hat{a}}\,\bm{\hat{b}} =
\frac{1}{2} \,
\begin{bmatrix}
+1 & -1\\
-1 & +1
\end{bmatrix}, \quad \text{and}\quad
\bm{\hat{b}}\,\bm{\hat{a}} =
\frac{1}{2} \,
\begin{bmatrix}
+1 & +1\\
+1 & +1
\end{bmatrix},
\]
in addition to $\bm{\hat{a}}$, $\bm{\hat{b}}$, and $\bm{\hat{e}}$.

\subsection{Closed-Form Solution of the Fundamental Equation (\ref{fundamental_equation_eom_3})}
By using $\bm{\mathcal{L}} = \sqrt{\bm{\mathcal{D}}} \, \bm{\mathcal{H}}$, 
the expansion of $\bm{\mathcal{\Hat{H}}}^n$ is expressed as 
\begin{align}
\bm{\mathcal{\Hat{H}}}^2 &= \left(\bm{\mathcal{H}} \otimes \bm{\hat{a}} + \sqrt{\bm{\mathcal{D}}} \otimes \bm{\hat{b}}\right)\left( \bm{\mathcal{H}} \otimes \bm{\hat{a}} + \sqrt{\bm{\mathcal{D}}} \otimes \bm{\hat{b}}\right)
= \bm{\mathcal{H}} \, \sqrt{\bm{\mathcal{D}}} \otimes \bm{\hat{a}} \bm{\hat{b}}+ \bm{\mathcal{L}} \otimes \bm{\hat{b}} \bm{\hat{a}}, 
\notag\\
\bm{\mathcal{\Hat{H}}}^3 &= \left( \bm{\mathcal{H}} \sqrt{\bm{\mathcal{D}}} \otimes \bm{\hat{a}} \bm{\hat{b}}+  \bm{\mathcal{L}} \otimes \bm{\hat{b}} \bm{\hat{a}}\right)\left( \bm{\mathcal{H}} \otimes \bm{\hat{a}} +  \sqrt{\bm{\mathcal{D}}} \otimes\bm{\hat{b}}\right)
= \bm{\mathcal{H}}\bm{\mathcal{L}} \otimes \bm{\hat{a}} +  \bm{\mathcal{L}} \,\sqrt{\bm{\mathcal{D}}} \otimes \bm{\hat{b}},
\notag\\
\bm{\mathcal{\Hat{H}}}^4 &= \left(\bm{\mathcal{H}} \bm{\mathcal{L}} \otimes \bm{\hat{a}} +  \bm{\mathcal{L}}\sqrt{\bm{\mathcal{D}}} \otimes \bm{\hat{b}}\right)\left( \bm{\mathcal{H}} \otimes \bm{\hat{a}} +  \sqrt{\bm{\mathcal{D}}} \otimes \bm{\hat{b}}\right)
= \bm{\mathcal{H}}\bm{\mathcal{L}} \sqrt{\bm{\mathcal{D}}} \otimes \bm{\hat{a}}\bm{\hat{b}}+ \bm{\mathcal{L}}^2 \otimes \bm{\hat{b}}\bm{\hat{a}},
\notag\\
\bm{\mathcal{\Hat{H}}}^5 &= \left( \bm{\mathcal{H}} \bm{\mathcal{L}}\sqrt{\bm{\mathcal{D}}} \otimes \bm{\hat{a}} \bm{\hat{b}} + \bm{\mathcal{L}}^2 \otimes \bm{\hat{b}}\bm{\hat{a}}\right)\left( \bm{\mathcal{H}} \otimes \bm{\hat{a}} +  \sqrt{\bm{\mathcal{D}}} \otimes \bm{\hat{b}}\right)
= \bm{\mathcal{H}}\bm{\mathcal{L}}^2 \otimes \bm{\hat{a}} +   \bm{\mathcal{L}}^2 \sqrt{\bm{\mathcal{D}}} \otimes \bm{\hat{b}},
\notag\\
\bm{\mathcal{\Hat{H}}}^6 &= \bm{\mathcal{H}} \bm{\mathcal{L}}^2 \sqrt{\bm{\mathcal{D}}} \otimes \bm{\hat{a}} \bm{\hat{b}} + \bm{\mathcal{L}}^3 \otimes \bm{\hat{b}} \bm{\hat{a}},
\notag\\
\bm{\mathcal{\Hat{H}}}^7 &= \bm{\mathcal{H}} \bm{\mathcal{L}}^3 \otimes \bm{\hat{a}}+   \bm{\mathcal{L}}^3 \sqrt{\bm{\mathcal{D}}} \otimes \bm{\hat{b}}. 
\notag
\end{align}
In general, by using $\bm{\mathcal{H}} = \sqrt{\bm{\mathcal{D}}^{-1}}\,\bm{\mathcal{L}}$, we easily have 
\begin{align}
\bm{\mathcal{\Hat{H}}}^{2k} &= \sqrt{\bm{\mathcal{D}}^{-1}} \bm{\mathcal{L}}^k  \sqrt{\bm{\mathcal{D}}}\otimes \bm{\hat{a}}\bm{\hat{b}}+ \bm{\mathcal{L}}^k \otimes \bm{\hat{b}} \bm{\hat{a}}, 
\label{H^even}\\
\bm{\mathcal{\Hat{H}}}^{2k+1} &= \bm{\mathcal{H}} \bm{\mathcal{L}}^k \otimes \bm{\hat{a}} +   \bm{\mathcal{L}}^k \sqrt{\bm{\mathcal{D}}} \otimes \bm{\hat{b}}, 
\label{H^odd}
\end{align}
for $k \ge 0$. 
Using these relations, we  can express the solution (\ref{solution_formsl}) of the fundamental equation~(\ref{fundamental_equation_eom_3}) in closed-form. 
We substitute (\ref{H^odd}) and (\ref{H^even}) into 
\begin{align}
\bm{\Hat{x}}(t) &= \exp\!\left(-\ii\bm{\mathcal{\Hat{H}}}\,t\right)  \bm{\Hat{x}}(0) \notag\\
&= \left(\cos(\bm{\mathcal{\Hat{H}}}\,t) -\ii\,\sin(\bm{\mathcal{\Hat{H}}}\,t)\right) \bm{\Hat{x}}(0)
\notag\\
&= \left(\sum_{k=0}^\infty (-1)^k \,\frac{\bm{(\mathcal{\Hat{H}}}\,t)^{2k}}{(2k)!} 
- \ii \, \sum_{k=0}^\infty (-1)^k \, \frac{\bm{(\mathcal{\Hat{H}}}\,t)^{2k+1}}{(2k+1)!}\right) \bm{\Hat{x}}(0). 
\notag
\end{align}
Both infinite sums are expressed as trigonometric functions as
\begin{align*}
\cos(\bm{\mathcal{\Hat{H}}}\,t) &= \sum_{k=0}^\infty (-1)^k \, \frac{\bm{(\mathcal{\Hat{H}}}\,t)^{2k}}{(2k)!} \notag\\
  &=\left[\sqrt{\bm{\mathcal{D}}^{-1}} \, \left(\bm{I} -  \bm{\mathcal{L}}\,\frac{t^{2}}{2!}  + \bm{\mathcal{L}}^{2} \,\frac{t^{4}}{4!} - \cdots \right)  \sqrt{\bm{\mathcal{D}}}\right]\otimes (\bm{\hat{a}}\,\bm{\hat{b}})
+\left[\left(\bm{I}-\bm{\mathcal{L}}\,\frac{t^{2}}{2!} \, +\bm{\mathcal{L}}^{2}\,\frac{t^{4}}{4!} -\cdots \right)\right]\otimes (\bm{\hat{b}}\,\bm{\hat{a}})
  \notag\\
  &= \left[\sqrt{\bm{\mathcal{D}}^{-1}} \bm{P} \left(\bm{I} - \bm{\Omega}^{2}\,\frac{t^{2}}{2!}  +  \bm{\Omega}^{4}\,\frac{t^{4}}{4!} - \cdots \right) \bm{P}^{-1} \sqrt{\bm{\mathcal{D}}}\right]\otimes (\bm{\hat{a}}\,\bm{\hat{b}})
+\left[\bm{P} \, \left(\bm{I}-\bm{\Omega}^{2}\,\frac{t^{2}}{2!} \, +\bm{\Omega}^{4}\,\frac{t^{4}}{4!}-\cdots \right) \, \bm{P}^{-1}\right]\otimes (\bm{\hat{b}}\,\bm{\hat{a}})
  \notag\\
  &=\left(
   \sqrt{\bm{\mathcal{D}}^{-1}} \, \bm{P}\,
  \cos(\bm{\Omega}\,t) \,
  \bm{P}^{-1} \, \sqrt{\bm{\mathcal{D}}}\right) \otimes (\bm{\hat{a}}\,\bm{\hat{b}})
  +\left(\bm{P} \, \cos(\bm{\Omega}\,t) \, \bm{P}^{- 1}\right) \otimes (\bm{\hat{b}}\,\bm{\hat{a}}),
\end{align*}
and
\begin{align*}
\sin(\bm{\mathcal{\Hat{H}}}\,t) &= \sum_{k=0}^\infty (-1)^k \, \frac{\bm{(\mathcal{\Hat{H}}}\,t)^{2k+1}}{(2k+1)!}
  \notag\\
  &= \left(\bm{\mathcal{H}}\,t - \bm{\mathcal{H}} \bm{\mathcal{L}}\,\frac{t^{3}}{3!} +\bm{\mathcal{H}} \bm{\mathcal{L}}^{2}\,\frac{t^{5}}{5!} -\cdots \right)\otimes \bm{\hat{a}}
+ \left(\sqrt{\bm{\mathcal{D}}}\,t - \bm{\mathcal{L}} \sqrt{\bm{\mathcal{D}}} \, \frac{t^{3}}{3!} + \bm{\mathcal{L}}^{2}  \sqrt{\bm{\mathcal{D}}}\,\frac{t^{5}}{5!} - \cdots  \right) \otimes \bm{\hat{b}}
  \notag\\
  &= \left[\sqrt{\bm{\mathcal{D}}^{-1}} \bm{P} \bm{\Omega}\left(\bm{\Omega}\,t - \bm{\Omega}^{3}\,\frac{t^{3}}{3!}  + \bm{\Omega}^{5} \, \frac{t^{5}}{5!} - \cdots  \right) \bm{P}^{-1}\right] \otimes \bm{\hat{a}}
+ \left[\bm{P}\,\bm{\mho} \left(\bm{\Omega}\,t - \bm{\Omega}^{3}\,\frac{t^{3}}{3!} + \bm{\Omega}^{5} \, \frac{t^{5}}{5!} - \cdots \right) \bm{P}^{-1} \sqrt{\bm{\mathcal{D}}}\right] \otimes \bm{\hat{b}}
  \notag\\
  &=
  \left(\sqrt{\bm{\mathcal{D}}^{-1}} \, \bm{P} \, \bm{\Omega} \, \sin(\bm{\Omega}\,t) \, \bm{P}^{-1}\right) \otimes \bm{\hat{a}}
  +\left(\bm{P} \,\bm{\mho} \, \sin(\bm{\Omega}\,t) \, \bm{P}^{-1} \, \sqrt{\bm{\mathcal{D}}}\right) \otimes \bm{\hat{b}},
\end{align*}
where 
\begin{align*}
\cos(\bm{\Omega}\,t) &= \mathrm{diag}\!\left(\cos(\omega_0\,t),\,\cos(\omega_1\,t),\,\dots,\,\cos(\omega_{n-1}\,t)\right),
\\
\sin(\bm{\Omega}\,t) &= \mathrm{diag}\!\left(\sin(\omega_0\,t),\,\sin(\omega_1\,t),\,\dots,\,\sin(\omega_{n-1}\,t)\right),
\\
\bm{\mho} &:= \text{diag}\!\left(0,\,1/\omega_1,\,\dots,\,1/\omega_{n-1}\right). 
\end{align*}
Therefore, the closed-form solution is given by
\begin{align}
\bm{\Hat{x}}(t) 
&= \Big(\sqrt{\bm{\mathcal{D}}^{-1}} \bm{P} \, \cos(\bm{\Omega}\,t) \, \bm{P}^{-1}  \sqrt{\bm{\mathcal{D}}} \otimes \bm{\hat{a}}\bm{\hat{b}}
 + \bm{P} \, \cos(\bm{\Omega}\,t) \, \bm{P}^{-1} \otimes \bm{\hat{b}}\bm{\hat{a}}\Big) \, \bm{\Hat{x}}(0) 
\notag\\
&\quad -\ii \, \Big(\sqrt{\bm{\mathcal{D}}^{-1}} \bm{P} \, \bm{\Omega}\,\sin(\bm{\Omega}\,t) \, \bm{P}^{-1} \otimes \bm{\hat{a}}
 + \bm{P} \,  \bm{\mho} \, \sin(\bm{\Omega}\,t) \, \bm{P}^{-1} \sqrt{\bm{\mathcal{D}}} \otimes \bm{\hat{b}} \Big) \, \bm{\Hat{x}}(0).
\label{eq:sol-F}
\end{align}

\section{General Solutions of the Wave Equation Derived from Fundamental Equations}
\label{sec.5}
The solution (\ref{eq:sol-F}) of the fermion-type fundamental equation (\ref{fundamental_equation_eom_3}) looks very different in form from the solution (\ref{solution-sqrt{L}}) of the boson-type fundamental equation (\ref{fundamental_equation_eom_two}).
In this section, we derive the solutions of the original wave equation \eqref{eom} from the closed-form solutions of the two different fundamental equations (\ref{fundamental_equation_eom_two}) and (\ref{fundamental_equation_eom_3}), and verify that both solutions are general solutions of the original wave equation.

\subsection{General Solutions of the Wave Equation}
As shown in Sec.~\ref{sec.3}, solutions of the fundamental equations (\ref{fundamental_equation_eom_two}) and (\ref{fundamental_equation_eom_3}) should be able to generate solutions of the original wave equation \eqref{eom}.
That is, for solutions $\bm{x}^+(t)$ and $\bm{x}^-(t)$ of the fundamental equations (\ref{fundamental_equation_eom_two}) and (\ref{fundamental_equation_eom_3}), the sum of them, $\bm{x}(t)=\bm{x}^+(t)+\bm{x}^-(t)$, should be able to give solutions of \eqref{eom}. 

In the following, the solution of the boson-type fundamental equation will be referred to as $\bm{x}^\pm_\text{b}(t)$, and the solution of the fermion-type fundamental equation will be referred to as $\bm{x}^\pm_\text{f}(t)$ in order to clarify which equation is the solution.
However, since the initial conditions are given in common, it will be referred as  $\bm{x}^\pm(0)$ in both equations. 

Let us consider the boson-type fundamental equation (\ref{fundamental_equation_eom_two}) first. 
We can denote its solution as 
\[
\bm{\hat{x}}_\mathrm{b}(t) = \bm{x}_\mathrm{b}^+(t) \otimes 
\begin{pmatrix}
1\\
0
\end{pmatrix}
+\bm{x}_\mathrm{b}^-(t) \otimes 
\begin{pmatrix}
0\\
1
\end{pmatrix}.
\] 
From the solutions (\ref{solution-x-pm}) of the boson-type fundamental equation (\ref{fundamental_equation_eom_two}), a solution of the original wave equation \eqref{eom} is obtained as $\bm{x}_\mathrm{b}(t):=\bm{x}_\mathrm{b}^+(t)+\bm{x}_\mathrm{b}^-(t)$, and it is expressed as 
\begin{align}
\bm{x}_\mathrm{b}(t) 
&= \bm{P} \, \exp(-\ii\,\bm{\Omega}\,t) \, \bm{P}^{-1} \, \bm{x}^+(0) 
+ \bm{P} \, \exp(+\ii\,\bm{\Omega}\,t) \, \bm{P}^{-1} \, \bm{x}^-(0), 
\label{solutionx_b}
\end{align}
for the initial condition $\bm{x}_\mathrm{b}(0)= \bm{x}^+(0) + \bm{x}^-(0)$. 

The second-order derivative of $\bm{x}_\mathrm{b}(t)$ with respect to $t$ gives
\begin{align}
\frac{\dd^2}{\dd t^2} \, \bm{x}_\mathrm{b}(t) 
&= \frac{\dd^2}{\dd t^2} \, \bm{P} \, \exp(-\ii\,\bm{\Omega}\,t) \, \bm{P}^{-1} \, \bm{x}^+(0) 
+ \frac{\dd^2}{\dd t^2} \, \bm{P} \, \exp(+\ii\,\bm{\Omega}\,t) \, \bm{P}^{-1} \, \bm{x}^-(0)
\notag\\
&= -\ii\,\frac{\dd}{\dd t} \, \bm{P} \,\bm{\Omega}\, \exp(-\ii\,\bm{\Omega}\,t) \, \bm{P}^{-1} \, \bm{x}^+(0) 
+\ii\, \frac{\dd}{\dd t} \, \bm{P} \,\bm{\Omega}\, \exp(+\ii\,\bm{\Omega}\,t) \, \bm{P}^{-1} \, \bm{x}^-(0)
\notag\\
&= -\bm{P}\, \bm{\Omega}^2\, \exp(-\ii\,\bm{\Omega}\,t) \, \bm{P}^{-1} \, \bm{x}^+(0) 
-\bm{P} \,\bm{\Omega}^2\, \exp(+\ii\,\bm{\Omega}\,t) \, \bm{P}^{-1} \, \bm{x}^-(0)
\notag\\
&= -\left(\bm{P}\,\bm{\Omega}^2 \bm{P}^{-1}\right) \left(\bm{P} \,\exp(-\ii\,\bm{\Omega}\,t) \, \bm{P}^{-1} \, \bm{x}^+(0) \right)
-\left(\bm{P} \,\bm{\Omega}^2 \bm{P}^{-1}\right) \left( \bm{P}\, \exp(+\ii\,\bm{\Omega}\,t) \, \bm{P}^{-1} \, \bm{x}^-(0) \right)
\notag\\
&= -\bm{\mathcal{L}} \, \bm{x}_\mathrm{b}(t). 
\end{align}
This means $\bm{x}_\mathrm{b}(t)$ is a solution of the original wave equation (\ref{eom}). 
Since the original wave equation \eqref{eom} is the second-order differential equation, solution \eqref{solutionx_b} that includes two arbitrary constants $\bm{x}^+(0)$ and $\bm{x}^-(0)$ as the initial conditions, is a general solution of the original wave equation \eqref{eom}. 

For comparison, we rewrite (\ref{solutionx_b}) in the following form: 
\begin{align}
\bm{x}_\mathrm{b}(t) &= \bm{P} \, \cos(\bm{\Omega}\,t) \, \bm{P}^{-1} \left(\bm{x}^+(0)+\bm{x}^-(0)\right)
-\ii \,\bm{P} \,  \sin(\bm{\Omega}\,t) \, \bm{P}^{-1} \left(\bm{x}^+(0)-\bm{x}^-(0)\right),
\label{solutionx_b2}
\end{align}

Next, we consider the fermion-type fundamental equation (\ref{fundamental_equation_eom_3}) and denote its solution as 
\[
\bm{\hat{x}}_\mathrm{f}(t) = \bm{x}_\mathrm{f}^+(t) \otimes 
\begin{pmatrix}
1\\
0
\end{pmatrix}
+\bm{x}_\mathrm{f}^-(t) \otimes 
\begin{pmatrix}
0\\
1
\end{pmatrix}.
\] 
Solution (\ref{eq:sol-F}) can be rewritten as separate entities for $\bm{x}_\mathrm{f}^+(t)$ and $\bm{x}_\mathrm{f}^-(t)$ as follows: 
\begin{align}
\bm{x}_\mathrm{f}^+(t) &= \sqrt{\bm{\mathcal{D}}^{-1}} \, \bm{P} \, \cos(\bm{\Omega}\,t) \, \bm{P}^{-1} \, \sqrt{\bm{\mathcal{D}}}\,\,\frac{\bm{x}^+(0)-\bm{x}^-(0)}{2}
+ \bm{P} \, \cos(\bm{\Omega}\,t) \, \bm{P}^{-1} \,\frac{\bm{x}^+(0)+\bm{x}^-(0)}{2}
\notag\\
&\quad {}-\ii \,\sqrt{\bm{\mathcal{D}}^{-1}} \, \bm{P} \, \bm{\Omega}\,\sin(\bm{\Omega}\,t) \, \bm{P}^{-1} \,\frac{\bm{x}^+(0)+\bm{x}^-(0)}{2}
-\ii \,\bm{P} \,  \bm{\mho} \, \sin(\bm{\Omega}\,t) \, \bm{P}^{-1} \sqrt{\bm{\mathcal{D}}} \,\,\frac{\bm{x}^+(0)-\bm{x}^-(0)}{2},
\\
\bm{x}_\mathrm{f}^-(t) &= -\sqrt{\bm{\mathcal{D}}^{-1}} \, \bm{P} \, \cos(\bm{\Omega}\,t) \, \bm{P}^{-1} \, \sqrt{\bm{\mathcal{D}}}\,\,\frac{\bm{x}^+(0)-\bm{x}^-(0)}{2}
+ \bm{P} \, \cos(\bm{\Omega}\,t) \, \bm{P}^{-1} \,\frac{\bm{x}^+(0)+\bm{x}^-(0)}{2}
\notag\\
&\quad {}+\ii \,\sqrt{\bm{\mathcal{D}}^{-1}} \, \bm{P} \, \bm{\Omega}\,\sin(\bm{\Omega}\,t) \, \bm{P}^{-1} \,\frac{\bm{x}^+(0)+\bm{x}^-(0)}{2}
-\ii \,\bm{P} \,  \bm{\mho} \, \sin(\bm{\Omega}\,t) \, \bm{P}^{-1} \sqrt{\bm{\mathcal{D}}} \,\,\frac{\bm{x}^+(0)-\bm{x}^-(0)}{2},
\end{align}
for the initial condition $\bm{x}_\mathrm{f}(0)= \bm{x}^+(0) + \bm{x}^-(0)$. 
The sum $\bm{x}_\mathrm{f}(t) := \bm{x}_\mathrm{f}^+(t) + \bm{x}_\mathrm{f}^-(t)$ gives 
\begin{align}
\bm{x}_\mathrm{f}(t) &= \bm{P} \, \cos(\bm{\Omega}\,t) \, \bm{P}^{-1} \left(\bm{x}^+(0)+\bm{x}^-(0)\right)
-\ii \,\bm{P} \,  \bm{\mho} \, \sin(\bm{\Omega}\,t) \, \bm{P}^{-1} \sqrt{\bm{\mathcal{D}}} \left(\bm{x}^+(0)-\bm{x}^-(0)\right).
\label{x_f(0)}
\end{align}

The second-order derivative of $\bm{x}_\mathrm{f}(t)$ with respect to $t$ gives
\begin{align}
\frac{\dd^2}{\dd t^2} \, \bm{x}_\mathrm{f}(t) 
&= \frac{\dd^2}{\dd t^2} \, \bm{P} \, \cos(\bm{\Omega}\,t) \, \bm{P}^{-1} \,(\bm{x}^+(0)+\bm{x}^-(0))
-\ii \, \frac{\dd^2}{\dd t^2} \, \bm{P} \,  \bm{\mho} \, \sin(\bm{\Omega}\,t) \, \bm{P}^{-1} \sqrt{\bm{\mathcal{D}}} \,(\bm{x}^+(0)-\bm{x}^-(0))
\notag\\
&= -\frac{\dd}{\dd t} \, \bm{P} \,\bm{\Omega}\, \sin(\bm{\Omega}\,t) \, \bm{P}^{-1} \,(\bm{x}^+(0)+\bm{x}^-(0))
-\ii \, \frac{\dd}{\dd t} \, \bm{P} \, \bm{\mho} \, \bm{\Omega} \, \cos(\bm{\Omega}\,t) \, \bm{P}^{-1} \sqrt{\bm{\mathcal{D}}} \,(\bm{x}^+(0)-\bm{x}^-(0))
\notag\\
&= -\bm{P} \,\bm{\Omega}^2\, \cos(\bm{\Omega}\,t) \, \bm{P}^{-1} \,(\bm{x}^+(0)+\bm{x}^-(0))
+\ii \, \bm{P} \,\bm{\Omega}\,\sin(\bm{\Omega}\,t) \, \bm{P}^{-1} \sqrt{\bm{\mathcal{D}}} \,(\bm{x}^+(0)-\bm{x}^-(0))
\notag\\
&= -\bm{P} \,\bm{\Omega}^2\,\bm{P}^{-1}\, \bm{P}\,\cos(\bm{\Omega}\,t) \, \bm{P}^{-1} \,(\bm{x}^+(0)+\bm{x}^-(0))
- \bm{P} \,\bm{\Omega}^2\,\bm{P}^{-1}\,(-\ii \, \bm{P}\,\bm{\mho}\,\sin(\bm{\Omega}\,t) \, \bm{P}^{-1}) \sqrt{\bm{\mathcal{D}}} 
\,(\bm{x}^+(0)-\bm{x}^-(0))
\notag\\
&= -\bm{\mathcal{L}} \, \bm{x}_\mathrm{f}(t). 
\end{align}
This means $\bm{x}_\mathrm{f}(t)$ is also a solution of the original wave equation (\ref{eom}). 
Since solution \eqref{x_f(0)} includes two arbitrary parameters $\bm{x}^+(0)$ and $\bm{x}^-(0)$ as the initial conditions, it is a general solution of the original wave equation \eqref{eom}. 

The solution $\bm{x}_\mathrm{b}(t)$ of (\ref{solutionx_b}) (or equivalently (\ref{solutionx_b2})) derived from the boson-type fundamental equation (\ref{fundamental_equation_eom_two}) can be also derived directly by solving the original wave equation (\ref{eom}). 
Of course, it is a general solution of the original wave equation (\ref{eom}).
On the other hand, the solution $\bm{x}_\mathrm{f}(t)$ of (\ref{x_f(0)}) is derived from the fermion-type fundamental equation \eqref{fundamental_equation_eom_3}, and it is also a solution of the original wave equation (\ref{eom}). 
The solution $\bm{x}_\mathrm{f}(t)$ is different, at least in form, from the well-known solution \eqref{solutionx_b2}, and it is a new solution that cannot be derived directly by solving the original wave equation (\ref{eom}). 
However, interestingly enough, since the solution (\ref{x_f(0)}) also includes two arbitrary initial conditions, $\bm{x}^+(0)$ and $\bm{x}^-(0)$, it is a general solution of the original wave equation (\ref{eom}). 
That is, both solutions (\ref{solutionx_b2}) and (\ref{x_f(0)}) give the same set of solutions as the general solution, although the actual solutions for the given initial condition are different. 
Such difference implies the importance of the fermion-type fundamental equation (\ref{fundamental_equation_eom_3}) because it is more suitable for describing user dynamics in OSNs. 
We can expect that the user dynamics described by the fermion-type fundamental equation (\ref{fundamental_equation_eom_3}) include unknown characteristics. 

\section{Characteristics of the Fermion-type Solution}
\label{sec.6}
This section compares the solution of the wave equation derived from the boson-type fundamental equation $\bm{x}_\mathrm {b}(t)$ with the solution of the wave equation derived from the fermion-type fundamental equation $\bm{x}_\mathrm{f}(t)$ and clarifies the characteristics of the solution of the fermion-type fundamental equation.
First, we reprint the solutions \eqref{solutionx_b2} and \eqref{x_f(0)} to be compared as: 
\begin{align*}
\bm{x}_\mathrm{b}(t) &= \bm{P} \, \cos(\bm{\Omega}\,t) \, \bm{P}^{-1} \left(\bm{x}^+(0)+\bm{x}^-(0)\right)
-\ii \,\bm{P} \,  \sin(\bm{\Omega}\,t) \, \bm{P}^{-1} \left(\bm{x}^+(0)-\bm{x}^-(0)\right),
\notag\\
\bm{x}_\mathrm{f}(t) &= \bm{P} \, \cos(\bm{\Omega}\,t) \, \bm{P}^{-1} \left(\bm{x}^+(0)+\bm{x}^-(0)\right)
-\ii \,\bm{P} \,  \bm{\mho} \, \sin(\bm{\Omega}\,t) \, \bm{P}^{-1} \sqrt{\bm{\mathcal{D}}} \left(\bm{x}^+(0)-\bm{x}^-(0)\right).
\end{align*}
From the time derivatives of them, we also obtain the following: 
\begin{align}
\frac{\dd \bm{x}_\mathrm{b}(t)}{\dd t} &= -\bm{P} \, \bm{\Omega}\,\sin(\bm{\Omega}\,t) \, \bm{P}^{-1} \left(\bm{x}^+(0)+\bm{x}^-(0)\right)
-\ii \,\bm{P} \,\bm{\Omega}\, \cos(\bm{\Omega}\,t) \, \bm{P}^{-1} \left(\bm{x}^+(0)-\bm{x}^-(0)\right),
\label{v_b(0)}\\
\frac{\dd \bm{x}_\mathrm{f}(t)}{\dd t} &= -\bm{P} \, \bm{\Omega}\,\sin(\bm{\Omega}\,t) \, \bm{P}^{-1} \left(\bm{x}^+(0)+\bm{x}^-(0)\right)
-\ii \,\bm{P} \,  \bm{\mho} \, \bm{\Omega}\,\cos(\bm{\Omega}\,t) \, \bm{P}^{-1} \sqrt{\bm{\mathcal{D}}} \left(\bm{x}^+(0)-\bm{x}^-(0)\right).
\label{v_f(0)}
\end{align}

Let us compare the initial states of both fundamental equations.
The initial state is determined by the initial displacement and the initial velocity. 
The initial displacement is the displacement at time $t = 0$, that is, $\bm{x}_\mathrm{b}(0)$ for boson-type and $\bm{x}_\mathrm{f}(0)$ for fermion-type. 
The initial velocity is the velocity at time $t = 0$ given as 
\[
\bm{\dot{x}}_\mathrm{b}(0) := \left.\frac{\dd \bm{x}_\mathrm{b}(t)}{\dd t}\right|_{t=0}, 
\quad 
\bm{\dot{x}}_\mathrm{f}(0) := \left.\frac{\dd \bm{x}_\mathrm{f}(t)}{\dd t}\right|_{t=0}. 
\]
First, consider the initial displacement. 
By setting $t = 0$ for solutions \eqref{solutionx_b2} and \eqref{x_f(0)}, we obtain
\begin{align*}
\bm{x}_\mathrm{b}(0) &= \bm{x}^+(0)+\bm{x}^-(0),
\notag\\
\bm{x}_\mathrm{f}(0) &= \bm{x}^+(0)+\bm{x}^-(0).
\end{align*}
Therefore, there is no difference between the two solutions regarding the initial displacement.
Next, let us compare the initial velocity. 
By setting $t=0$ for \eqref{v_b(0)} and \eqref{v_f(0)}, we obtain
\begin{align}
\bm{\dot{x}}_\mathrm{b}(0) &= 
-\ii \,\bm{P} \, \bm{\Omega}\, \bm{P}^{-1} \left(\bm{x}^+(0)-\bm{x}^-(0)\right),
\label{init_veloc_b}\\
\bm{\dot{x}}_\mathrm{f}(0) &= 
-\ii \,\bm{P} \,  \bm{\mho} \, \bm{\Omega}\, \bm{P}^{-1} \sqrt{\bm{\mathcal{D}}} \left(\bm{x}^+(0)-\bm{x}^-(0)\right).
\label{init_veloc_f}
\end{align}
Since both \eqref{solutionx_b2} and \eqref{x_f(0)} are solutions of the original wave equation \eqref {eom}, if the initial velocity is $0$, both the solutions $\bm{x}_\mathrm{b}(t)$ and $\bm{x}_\mathrm{f}(t)$ are equal. 
Therefore, the difference between them becomes apparent when the initial velocity is not $0$.

To analyze the difference between the two solutions, let us give the components of the initial conditions $\bm{x}^+(0)$ and $\bm{x}^-(0)$ to be complex conjugates. 
If the components of $\bm{x}^+(0)$ and $\bm{x}^-(0)$ are expressed as
\begin{align*}
\bm{x}^+(t) = {}^t\!\left(x^+_1(t),\,x^+_2(t),\,\dots,\,x^+_n(t)\right),\\
\bm{x}^-(t) = {}^t\!\left(x^-_1(t),\,x^-_2(t),\,\dots,\,x^-_n(t)\right),
\end{align*}
let us give
\[
x^+_i(0) = a_i + \ii b_i, \quad x^-_i(0) = a_i - \ii b_i,
\]
where $a_i,\,b_i \in \mathbb{R}$ $(i=1,\,\dots,\,n)$.
When selected in this way, it becomes 
\[
x^+_i(0) + x^-_i(0) = 2a_i, \quad x^+_i(0) - x^-_i(0) = 2\ii b_i,
\]
and therefore we can treat the initial displacement and the initial velocity as real values.
Furthermore, the initial displacement is not needed to compare the solutions $\bm{x}_\mathrm{b}(t)$ and $\bm{x}_\mathrm{f}(t)$, so the comparison is performed under the initial condition where only the initial velocity given by $a_i = 0$. 

\begin{figure}[bt]
  \centering
\includegraphics[width=0.8\linewidth]{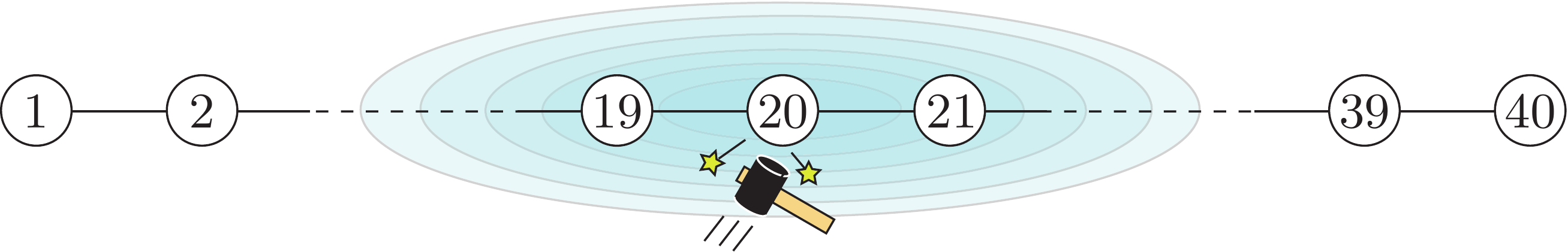}
\caption{1-dimensional network model and the initial condition}
    \label{fig:NW_model}
\end{figure}

In the following, we will compare and examine both solutions concretely by numerical experiments using a simple network model. 
Consider a one-dimensional undirected graph consisting of $ 40 $ nodes shown in Fig.~\ref{fig:NW_model}, and the weights of all the links are $1$. 
As an initial state, the initial displacement is given to be $0$ for all nodes, the initial velocity is given only to the node $20$, and the initial velocity of the other nodes is $0$.
As shown in Fig.~\ref{fig:NW_model}, this assumes a situation where the impact as the initial velocity is applied only to the node $ 20 $ at the time $ t = 0 $.
Specifically, the initial condition is given as
\[
a_i = 0, \qquad 
b_i = \left\{
\begin{array}{cc}
0     & \quad (i\not=20), \\
\displaystyle \frac{1}{2} & \quad (i=20),
\end{array}
\right.
\qquad 
x^\pm_i(0) = \left\{
\begin{array}{cc}
0     & \quad (i\not=20), \\
\displaystyle \pm\frac{1}{2} \,\ii & \quad (i=20).
\end{array}
\right.
\]

\begin{figure}[tb]
  \centering
\begin{tabular}{cc}
\includegraphics[width=0.48\linewidth]{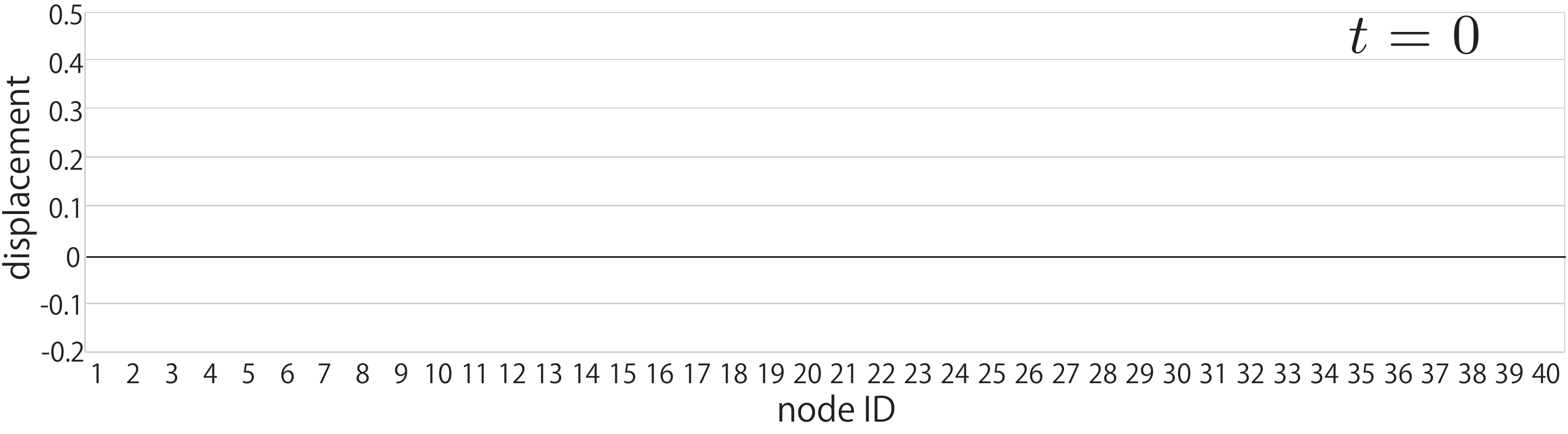}&
\includegraphics[width=0.48\linewidth]{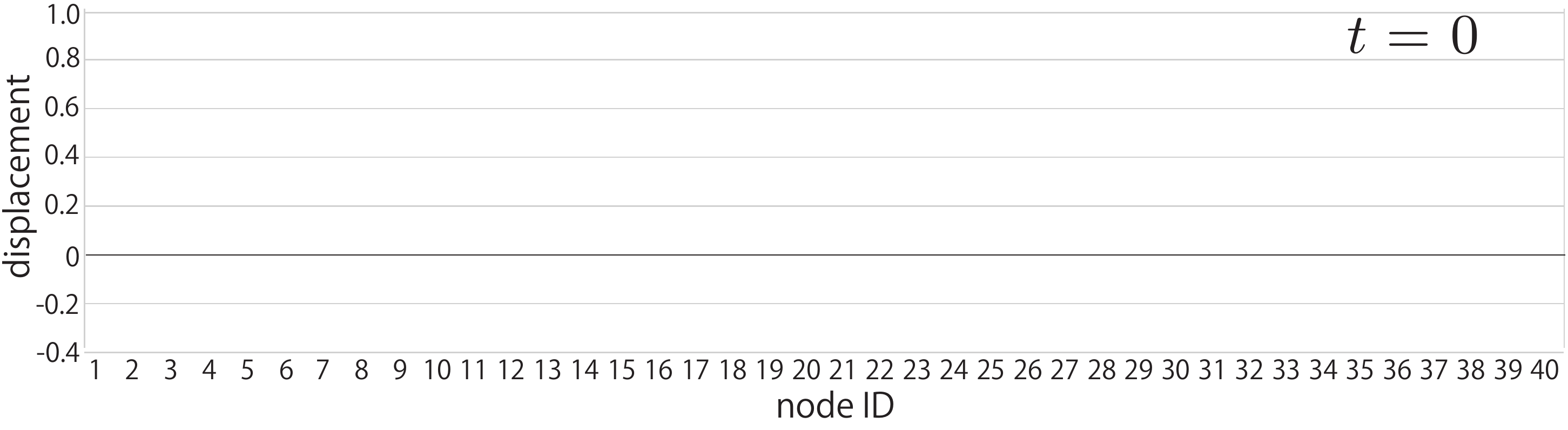}\\
\includegraphics[width=0.48\linewidth]{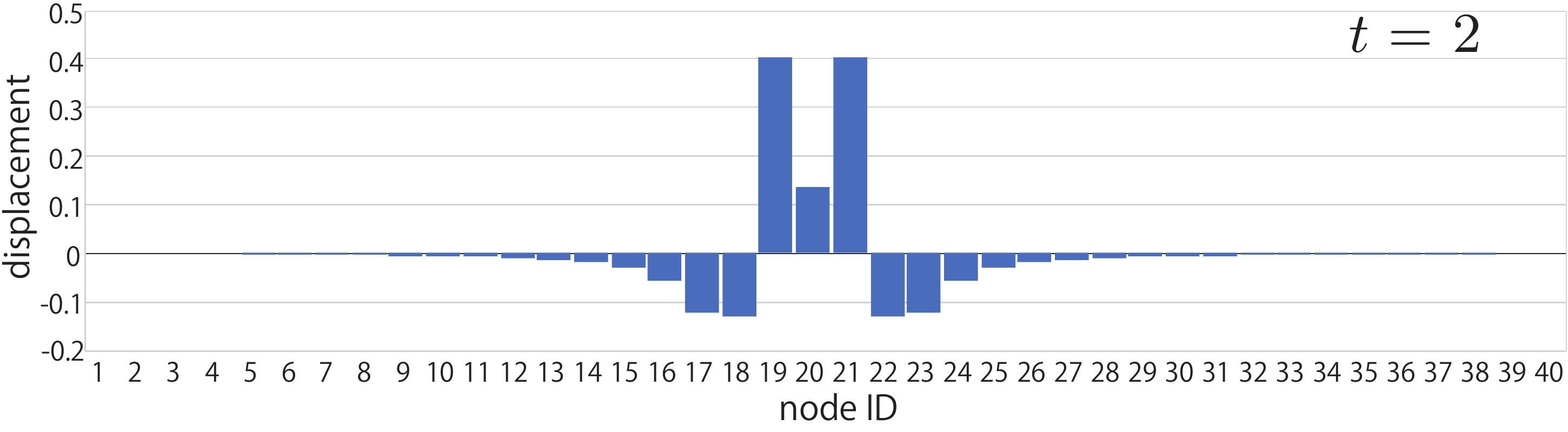}&
\includegraphics[width=0.48\linewidth]{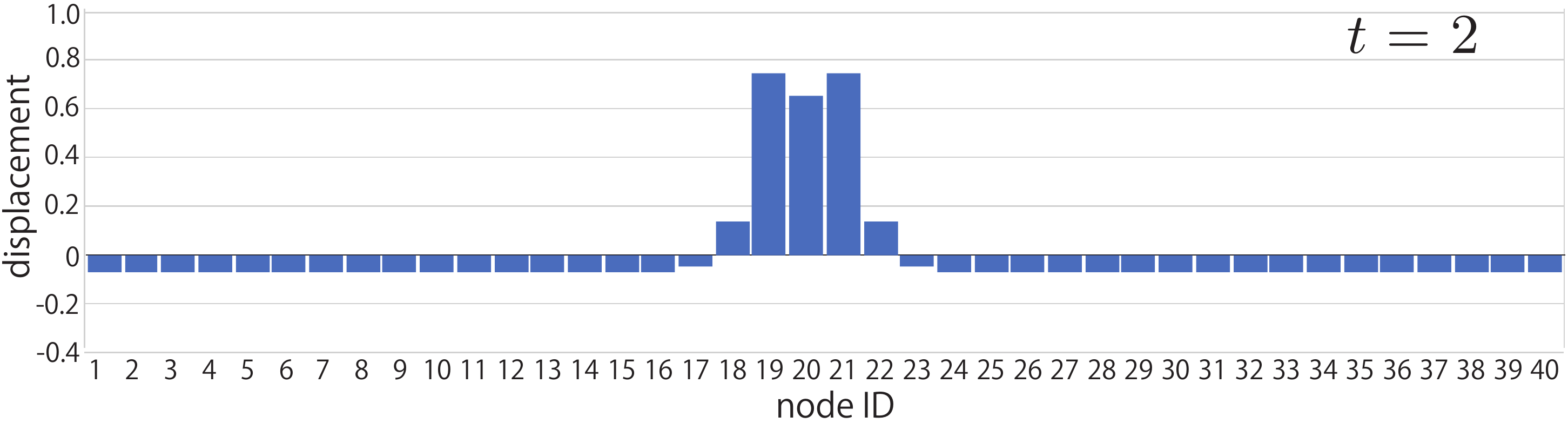}\\
\includegraphics[width=0.48\linewidth]{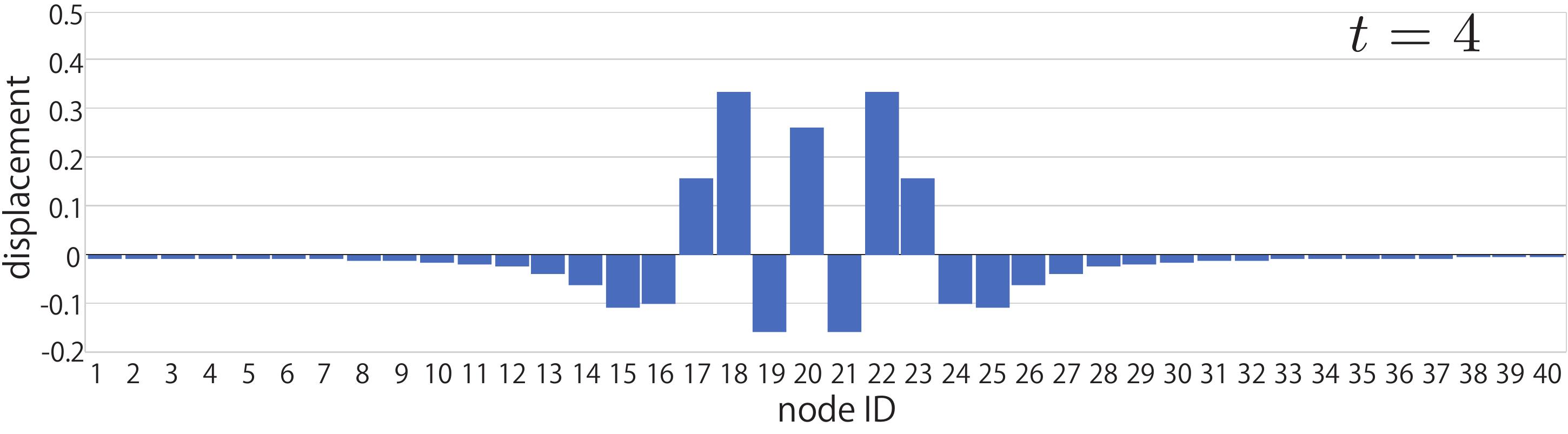}&
\includegraphics[width=0.48\linewidth]{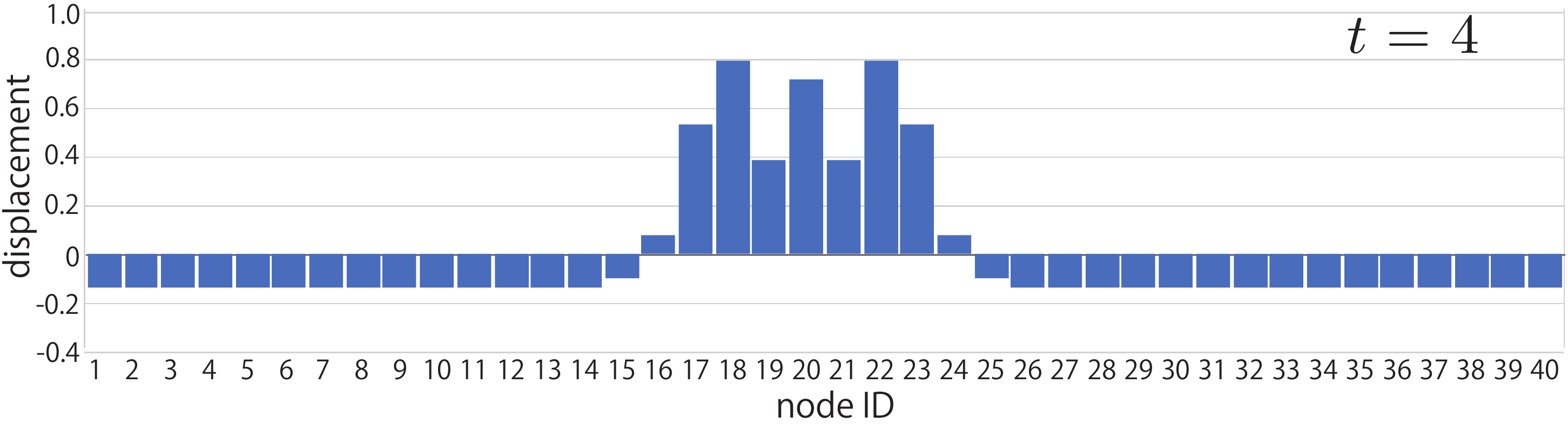}\\
\includegraphics[width=0.48\linewidth]{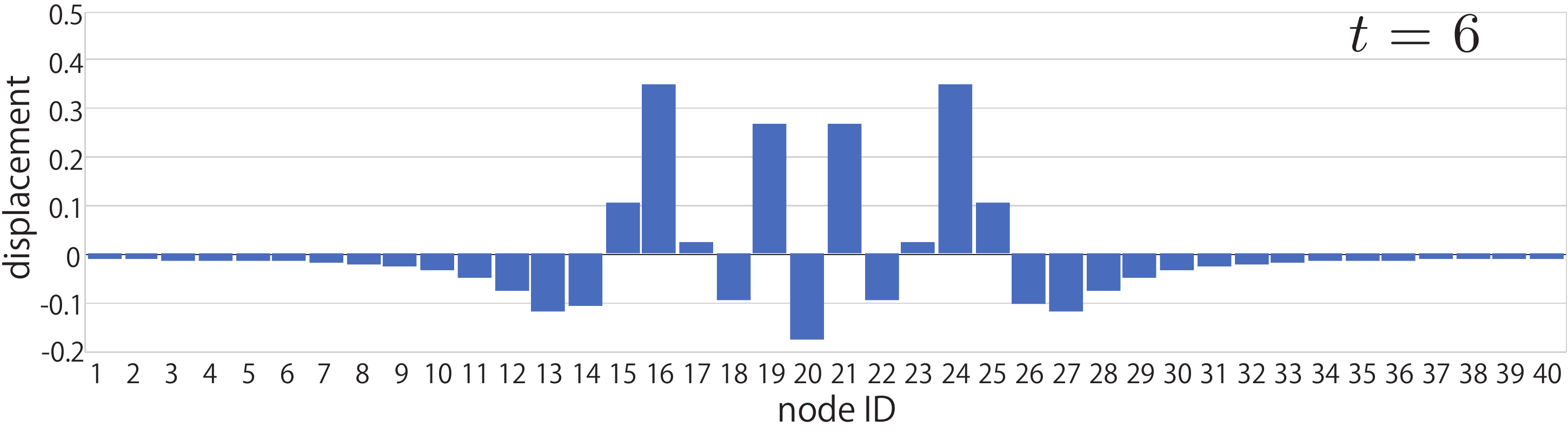}&
\includegraphics[width=0.48\linewidth]{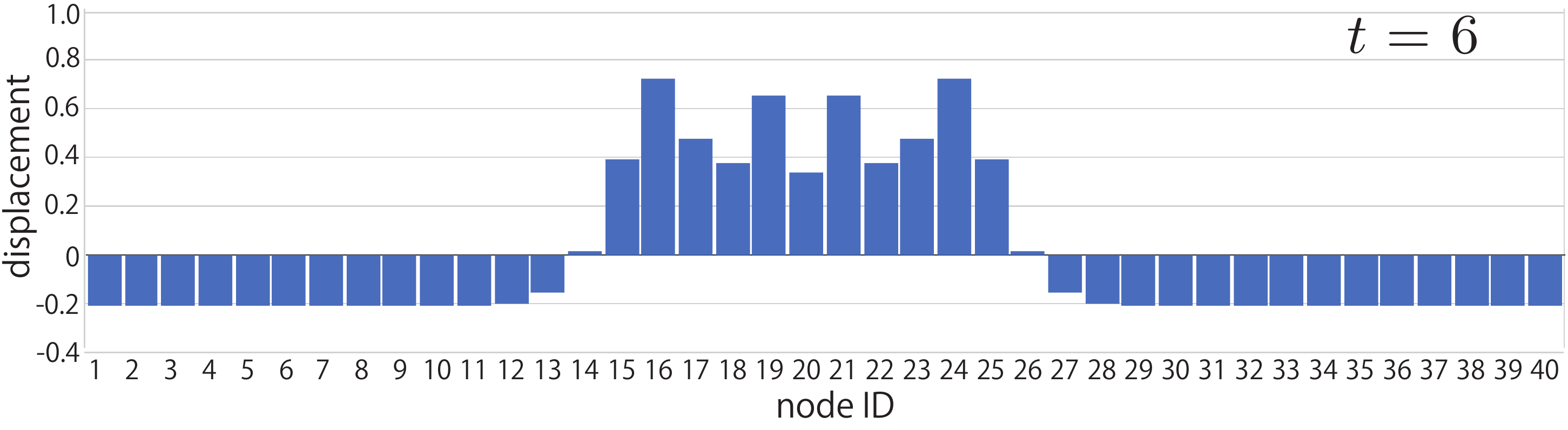}\\
\includegraphics[width=0.48\linewidth]{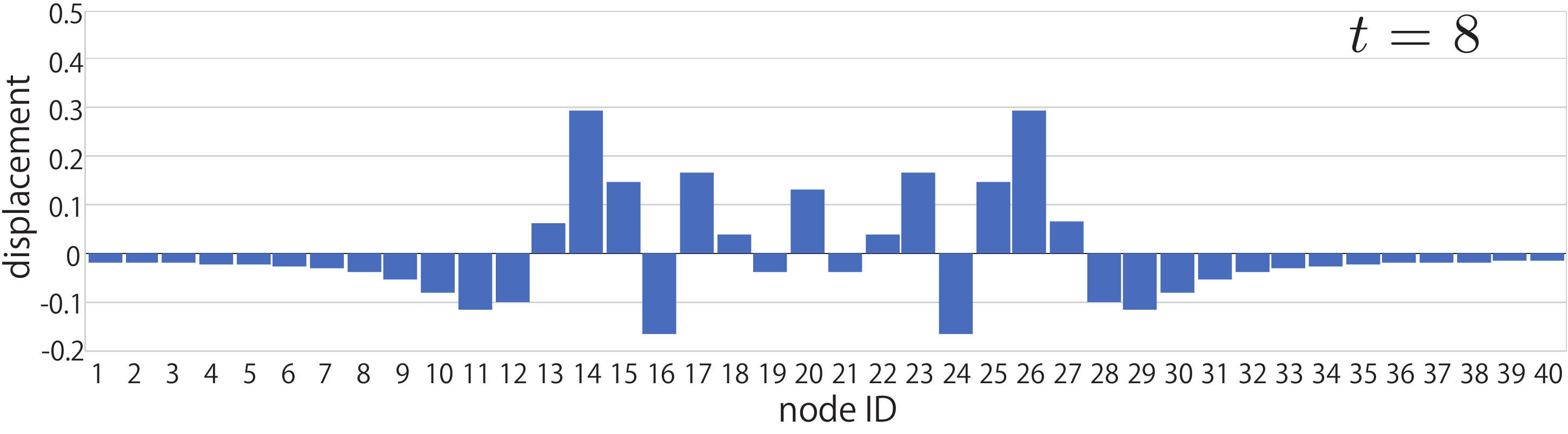}&
\includegraphics[width=0.48\linewidth]{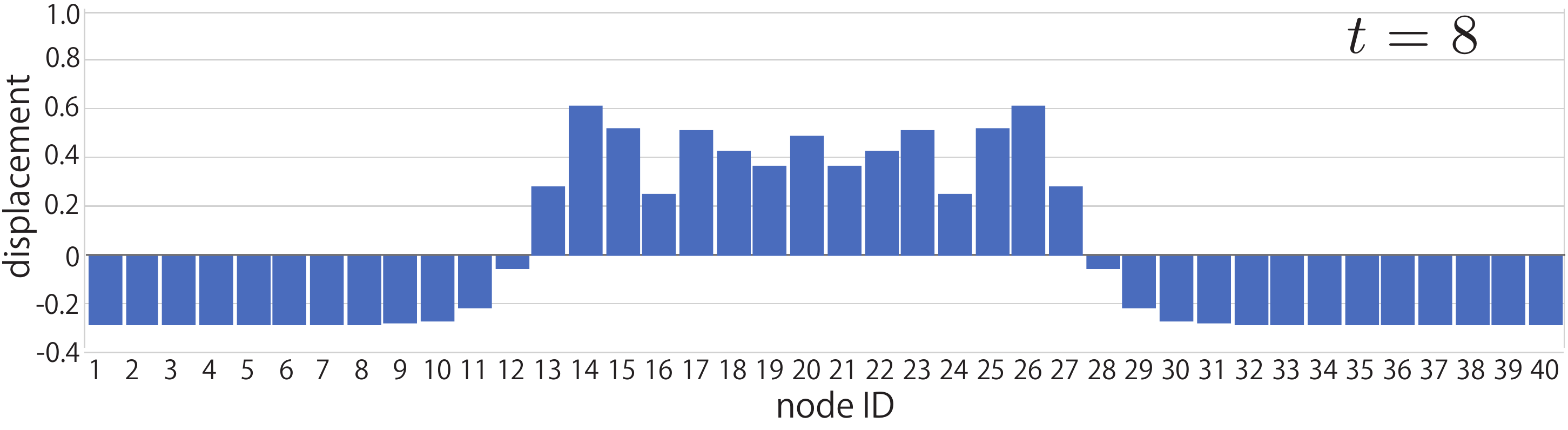}\\
\includegraphics[width=0.48\linewidth]{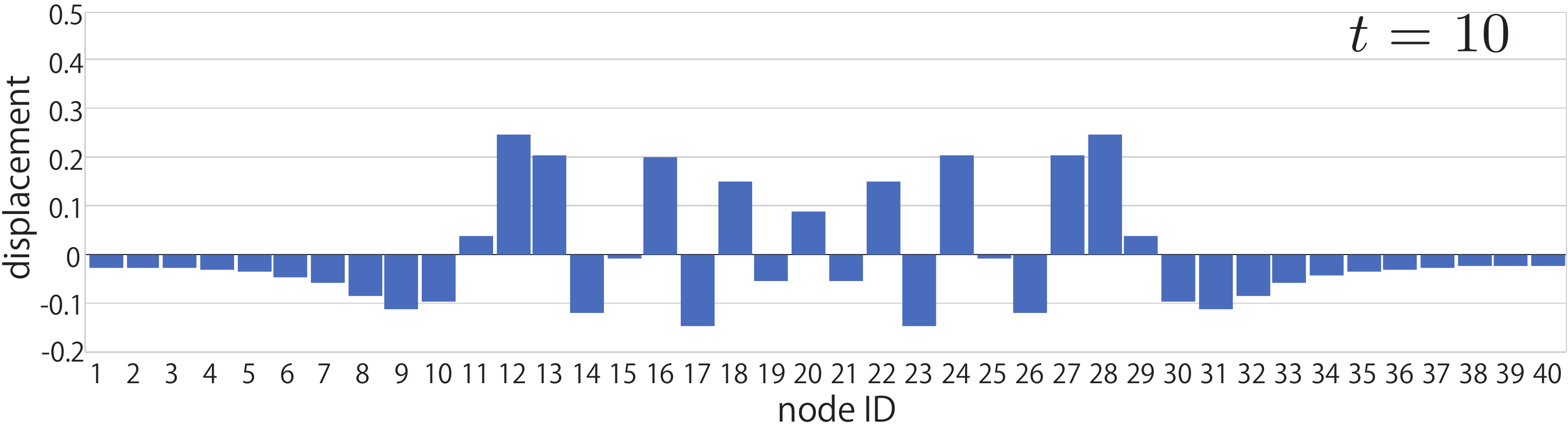}&
\includegraphics[width=0.48\linewidth]{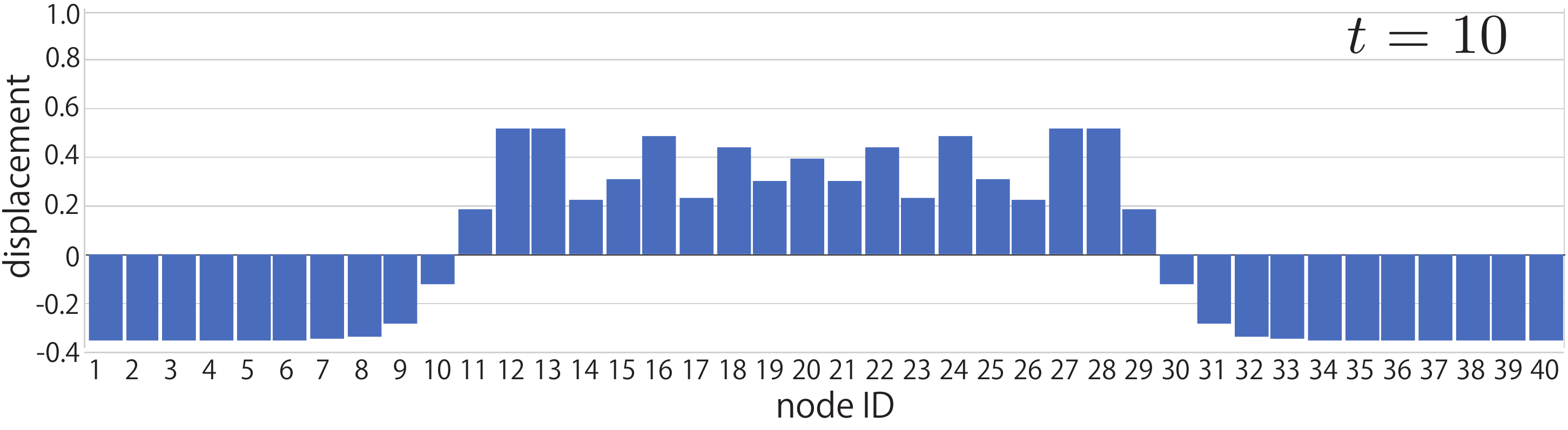}\\
(a) $\bm{x}_\mathrm{b}(t)$ for boson-type & (b) $\bm{x}_\mathrm{f}(t)$ for fermion-type
\end{tabular}
\caption{Temporal changes of the displacements of nodes for boson-type (left) and fermion-type (right) fundamental equations}
    \label{fig:displ}
\end{figure}

Figure~\ref{fig:displ} shows the displacement of each node in the above network model.
The left panel shows the boson-type solution $\bm{x}_\mathrm{b}(t)$, and the right panel shows the fermion-type solution $\bm{x}_\mathrm{f}(t)$.
In addition, the passage of time is shown in order from the graph above, the horizontal axis of each graph shows the node number of the one-dimensional network model, and the vertical axis shows the displacement of each node.

The following is what can be seen from these results.
First, at time $t = 0$, the displacements of all nodes are $0$ for both boson and fermion types.
This is because the initial displacement is given to be $ 0 $, which is common to both.
However, the behavior of both solutions is significantly different after that time.
From this, it can be understood that the boson-type and the fermion-type represent different solutions.

As a tendency common to both, it can be seen that the influence spreads symmetrically around the node $20$ that gave the initial velocity.
In this regard, the problem is causality.
Since the wave equation should be a framework for describing propagation at a finite speed in the first place, propagation at an infinite speed is not allowed. 
In this model, the weight of the link is given as $1$, so the propagation speed is $1$, and at time $t = 10$, propagation can only occur from the node $20$ to the left and right nodes of up to about $\pm 10$ nodes. 
Nevertheless, both appear to be propagating beyond that range.
Especially for the Fermi-type solution, displacement appears in the entire network from the initial stage, and at first glance, it seems that propagation occurs at an infinite speed.

\begin{figure}[tb]
  \centering
\begin{tabular}{cc}
\includegraphics[width=0.48\linewidth]{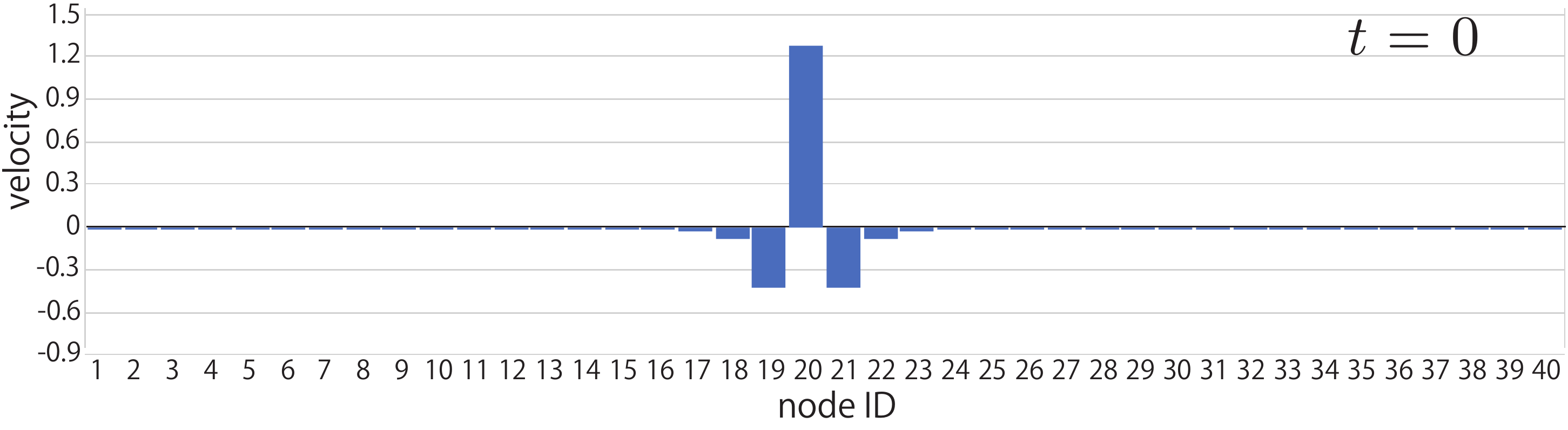}&
\includegraphics[width=0.48\linewidth]{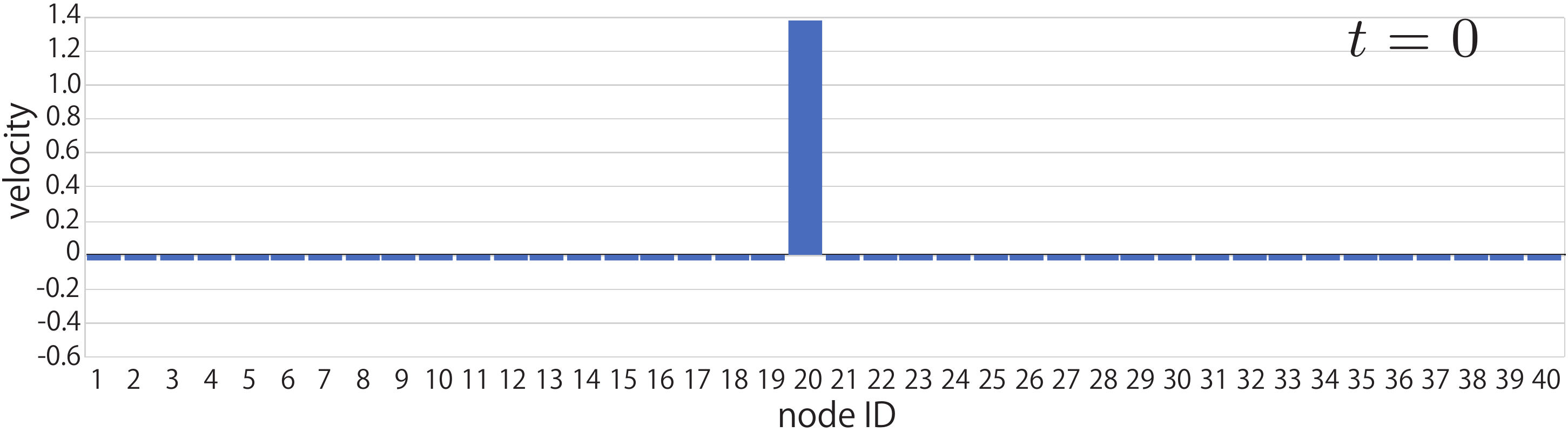}\\
\includegraphics[width=0.48\linewidth]{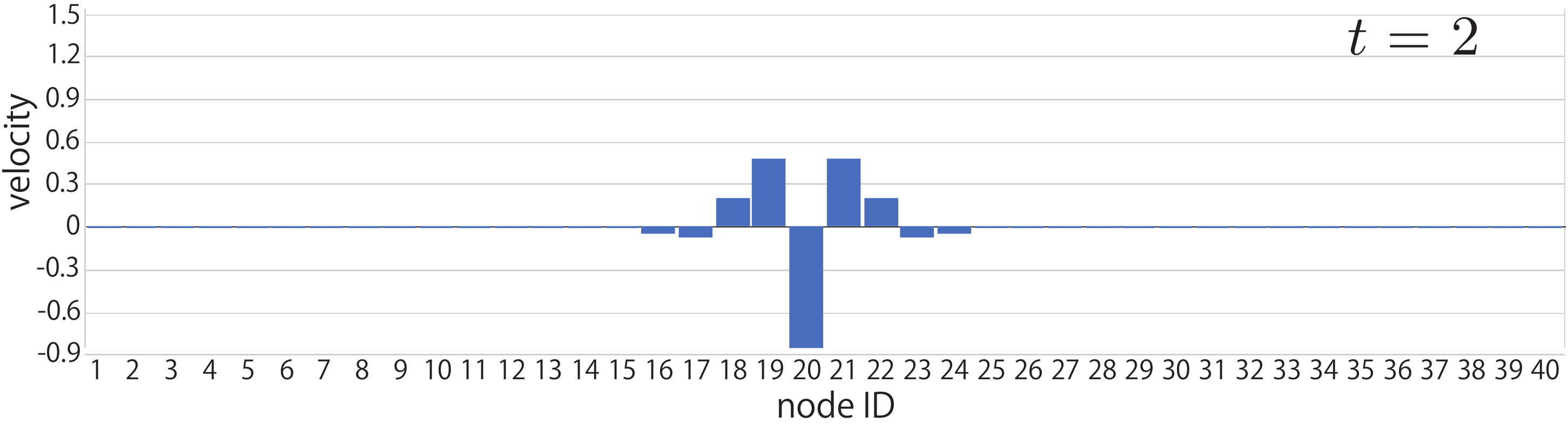}&
\includegraphics[width=0.48\linewidth]{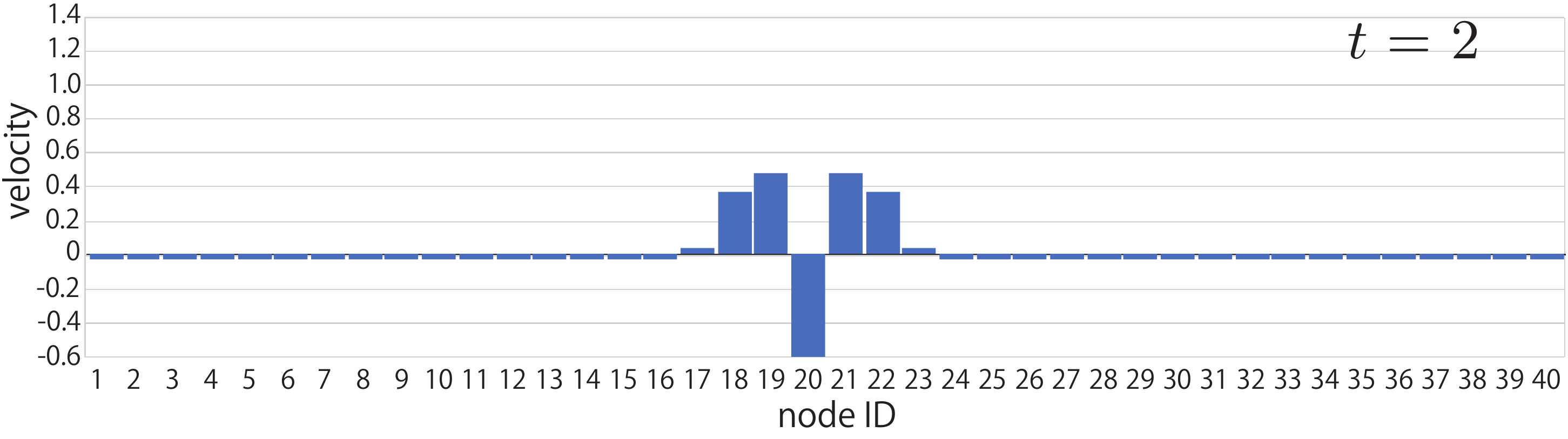}\\
\includegraphics[width=0.48\linewidth]{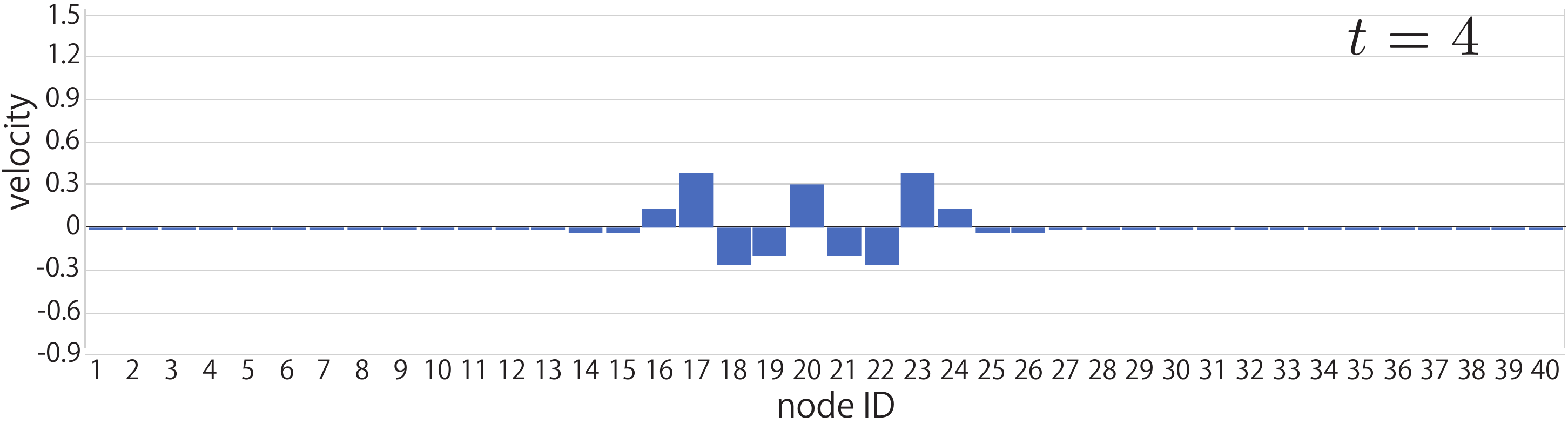}&
\includegraphics[width=0.48\linewidth]{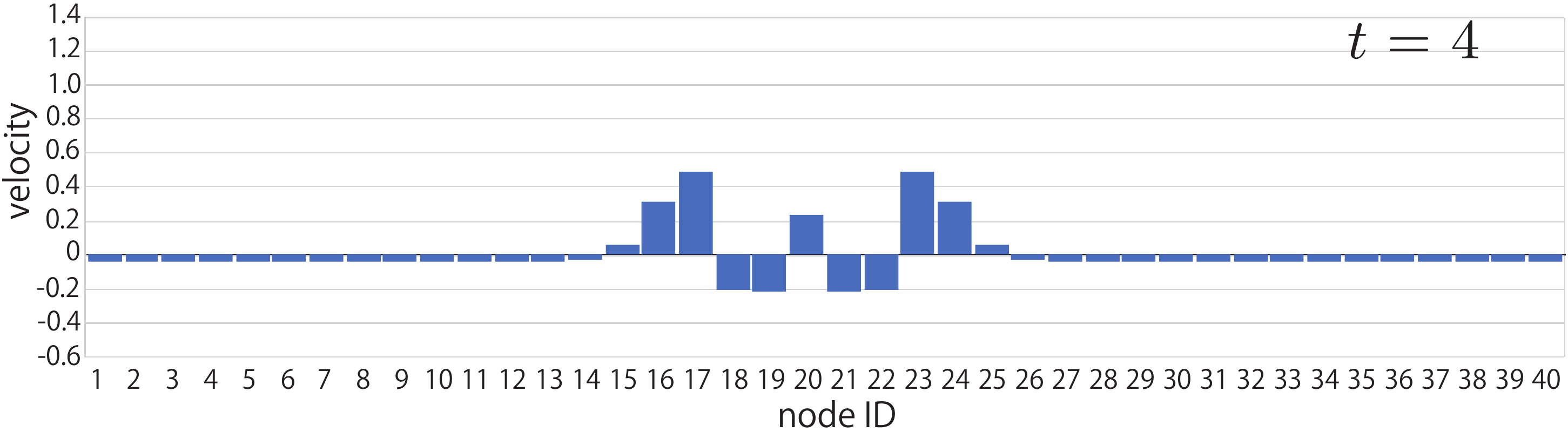}\\
\includegraphics[width=0.48\linewidth]{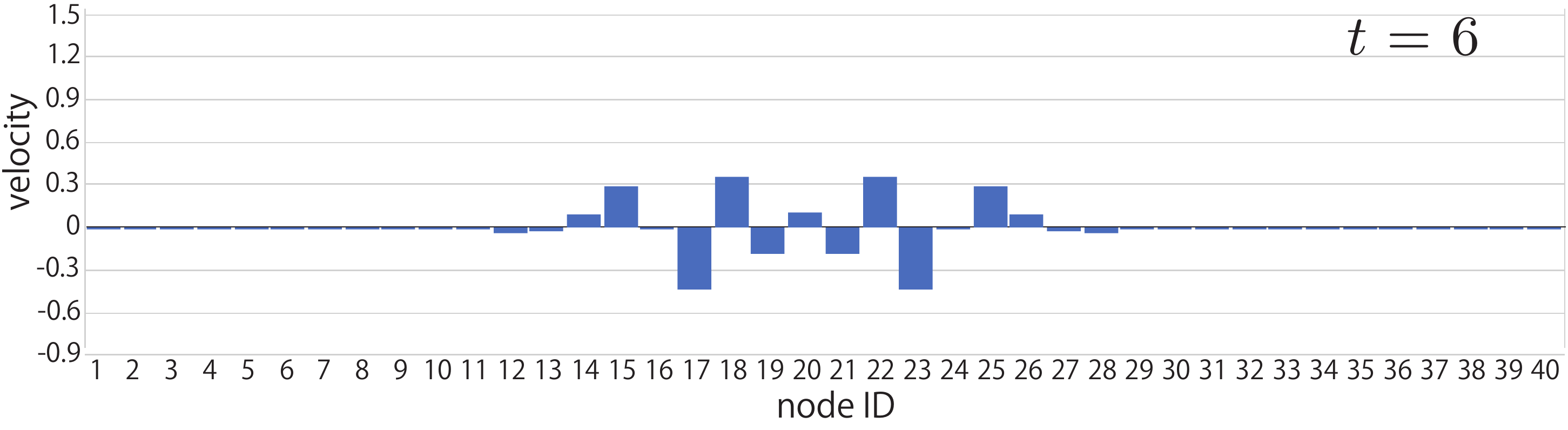}&
\includegraphics[width=0.48\linewidth]{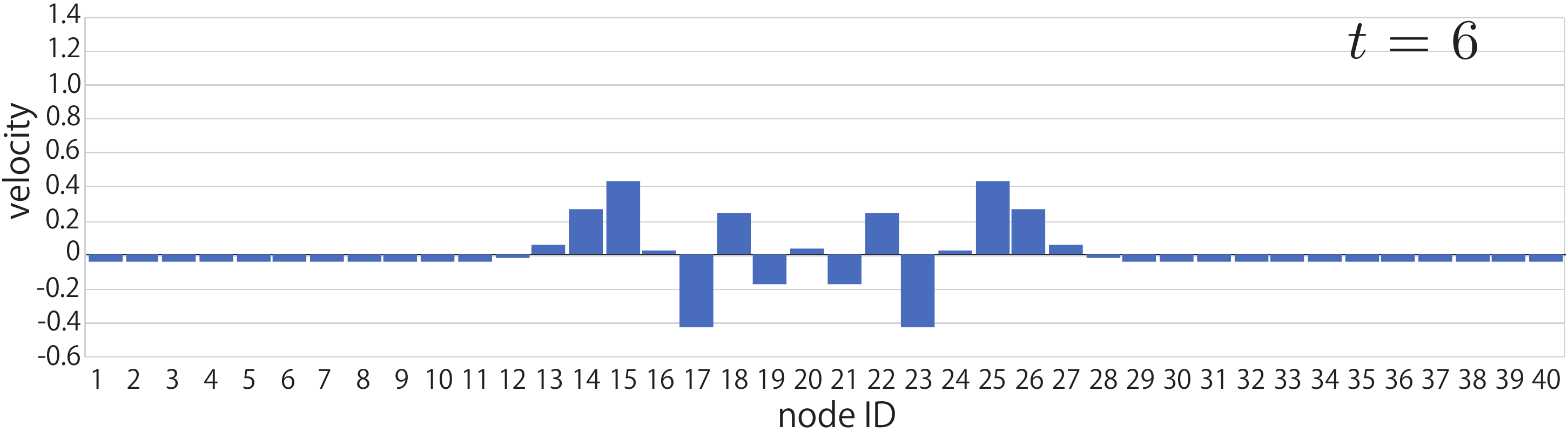}\\
\includegraphics[width=0.48\linewidth]{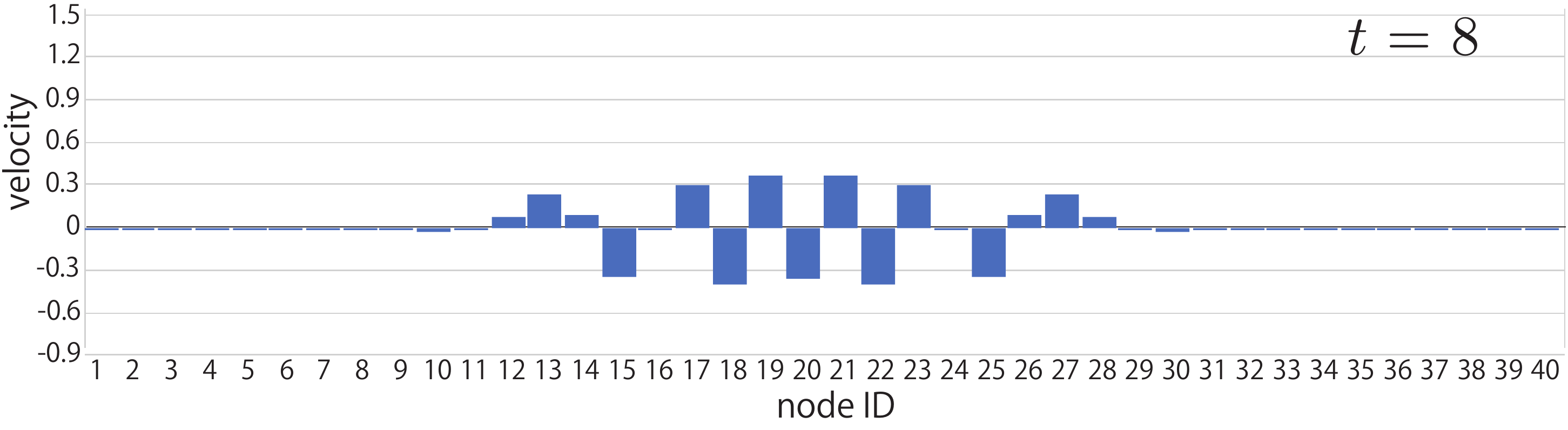}&
\includegraphics[width=0.48\linewidth]{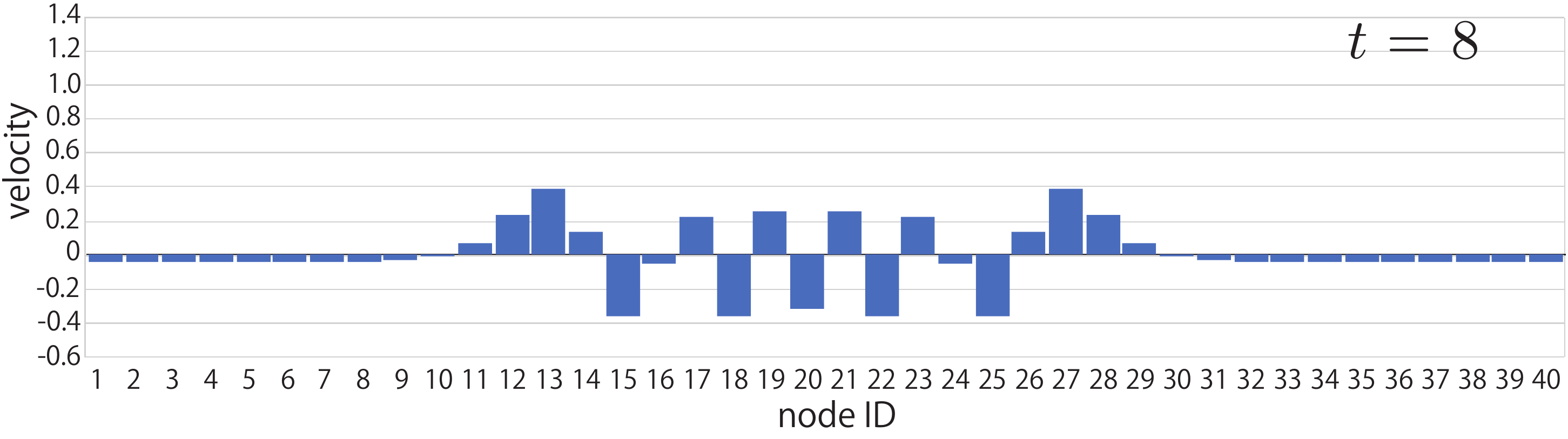}\\
\includegraphics[width=0.48\linewidth]{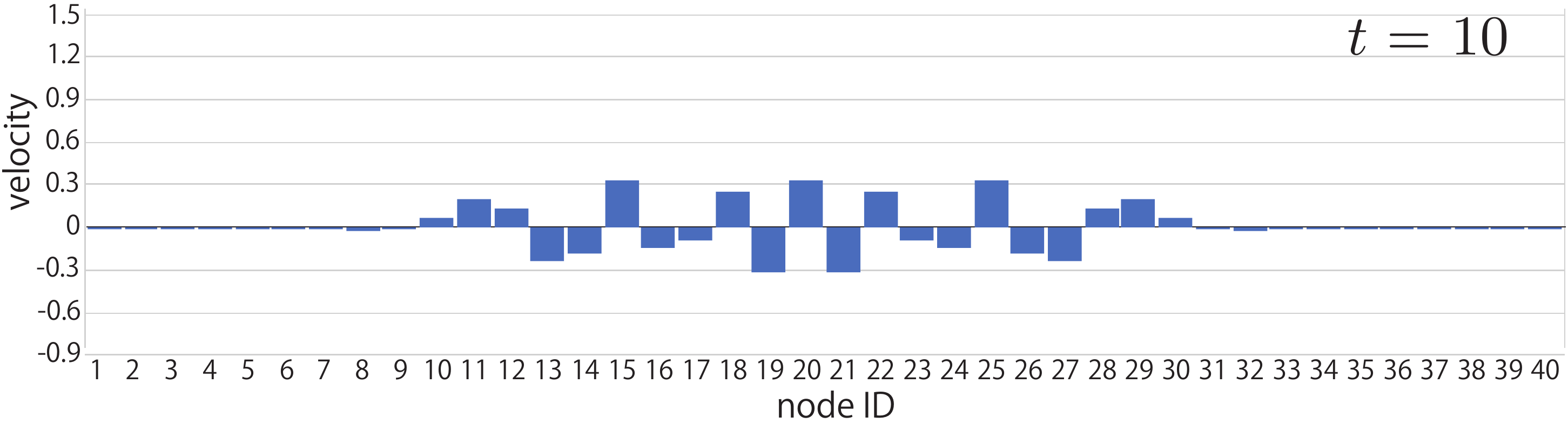}&
\includegraphics[width=0.48\linewidth]{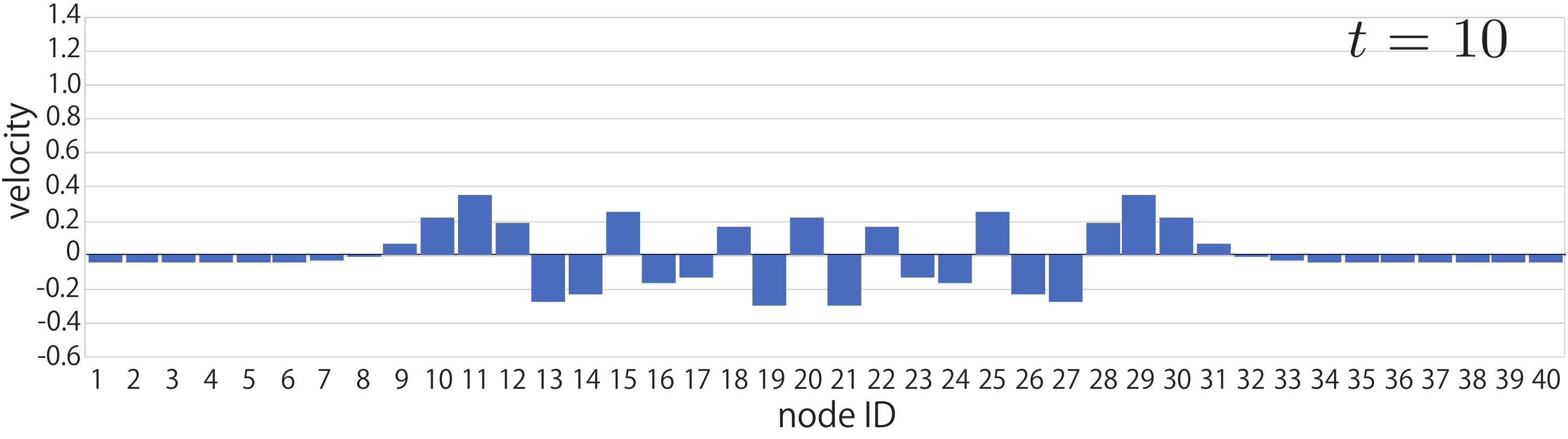}\\
(a) $\frac{\dd}{\dd t}\bm{x}_\mathrm{b}(t)$ for boson-type & (b) $\frac{\dd}{\dd t} \bm{x}_\mathrm{f}(t)$ for fermion-type
\end{tabular}
\caption{Temporal changes of the velocity of nodes for boson-type (left) and fermion-type (right) fundamental equations}
    \label{fig:veloc}
\end{figure}

To analyze why such violations of causality seem to occur, we examine the velocity changes at each node.
Figure~\ref{fig:veloc} shows the velocity of change of the displacement of each node.
The left panel shows the boson-type solution $\bm{x}_\mathrm{b}(t)$, and the right panel shows the Fermi type solution $\bm{x}_\mathrm{f}(t)$.
In addition, the passage of time is shown in order from the graph above, the horizontal axis of each graph shows the node number of the one-dimensional network model, and the vertical axis shows the velocity of change of displacement of each node.

What we can see immediately from this is that even though the initial velocity is given only to the node $20$, the initial state of the wave equation takes a value other than $0$ other than the node $20$.
For boson-type results, some nodes around node $20$ have non-zero initial velocity, but other nodes have an initial velocity of $0$.
In Fig.~\ref{fig:displ}, the reason why the boson-type displacement propagation occurs at a faster speed than expected is that the initial velocity is distributed around the node $20$ in this way.
Since the initial velocity already has a spatial spread at time $t = 0$, the displacement seems to propagate at an impossible velocity if we assume the influence propagates from the node $20$.

The cause of this phenomenon is as follows.
From \eqref{init_veloc_b} and 
\[
\sqrt{\bm{\mathcal{L}}} = \bm{P} \, \bm{\Omega}\, \bm{P}^{-1}, 
\]
the initial velocity of the boson-type is obtained as
\begin{align*}
\bm{\dot{x}}_\mathrm{b}(0) &= 
-\ii \,\sqrt{\bm{\mathcal{L}}} \left(\bm{x}^+(0)-\bm{x}^-(0)\right).
\end{align*}
Since $\sqrt{\bm{\mathcal{L}}}$ has components other than diagonal components, $\bm{\dot{x}}_\mathrm{b}(0)$ has non-zero components other than node $20$ even if $\bm{x}^+(0)-\bm{x}^-(0)$ has only one non-zero component corresponding to node $20$. 
Conversely, when the initial velocity $\bm{\dot{x}}_\mathrm{b}(0)$ is concentrated only on a specific node in the original wave equation, consider whether we can choose the corresponding $\bm{x}^+(0)$ and $\bm{x}^-(0)$. 
If there were an inverse matrix in $\bm{\mathcal{L}}$, we would obtain $\bm{x}^+(0)$ and $\bm{x}^-(0)$ corresponding to $\bm{\dot{x}}_\mathrm{b}(0)$ as
\begin{align*}
\bm{x}^+(0)-\bm{x}^-(0) &= \ii \,\sqrt{\bm{\mathcal{L}}^{-1}} \, \bm{\dot{x}}_\mathrm{b}(0). 
\end{align*}
Unfortunately, this is not possible because the Laplacian matrix does not have its inverse.
That is, no matter how we choose $\bm{x}^+(0)$ and $\bm{x}^-(0)$, an initial condition with only the node $20$ having velocity in the original wave equation \eqref{eom} cannot be created.

\begin{figure}[tb]
  \centering
\begin{tabular}{cc}
\includegraphics[width=0.48\linewidth]{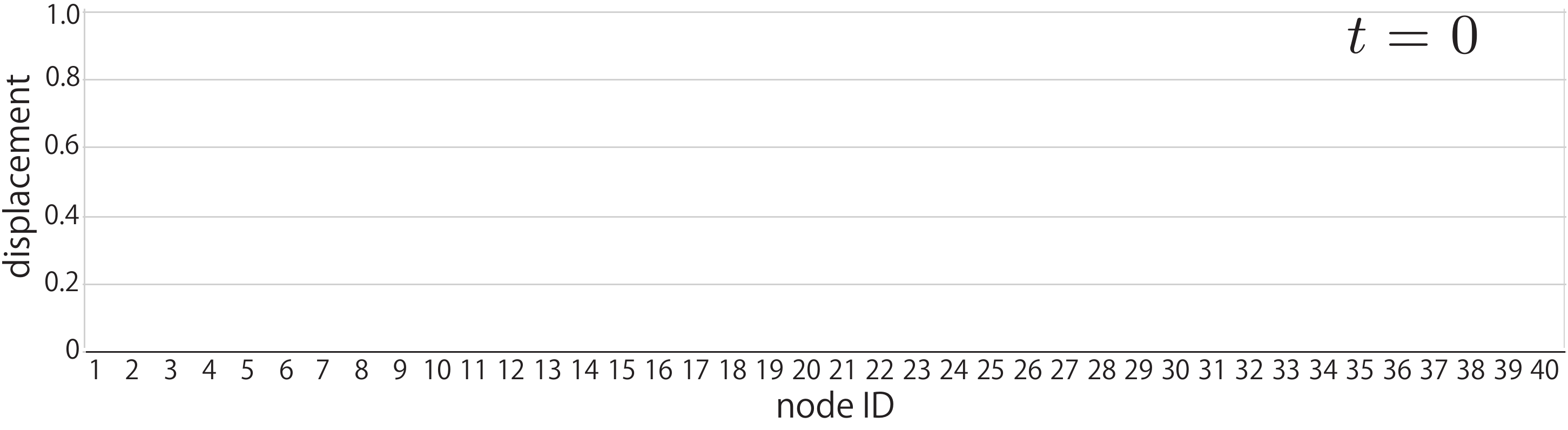}&
\includegraphics[width=0.48\linewidth]{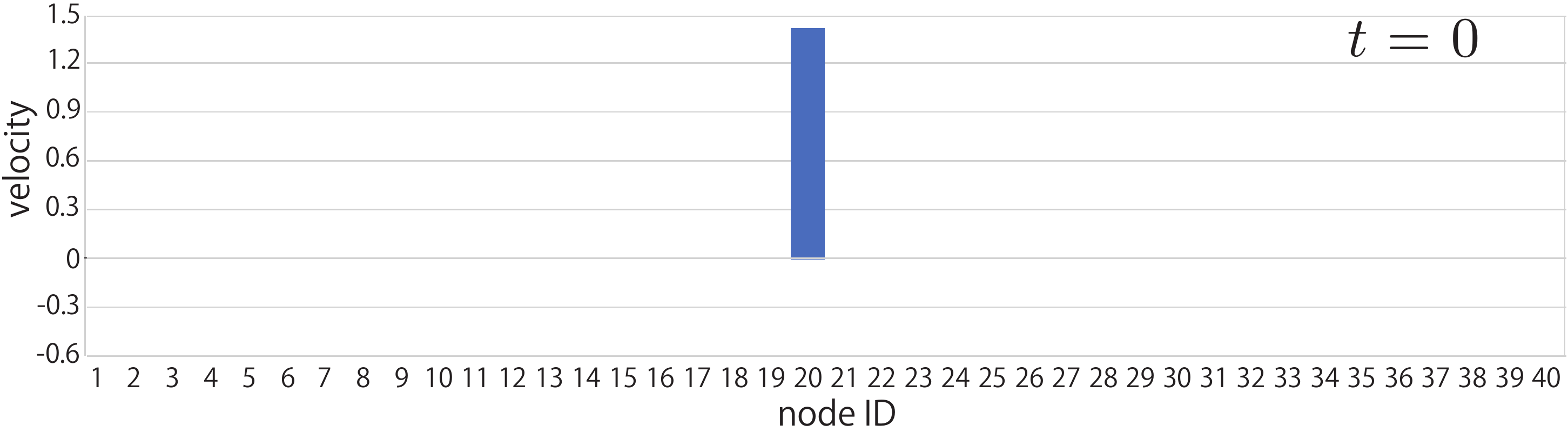}\\
\includegraphics[width=0.48\linewidth]{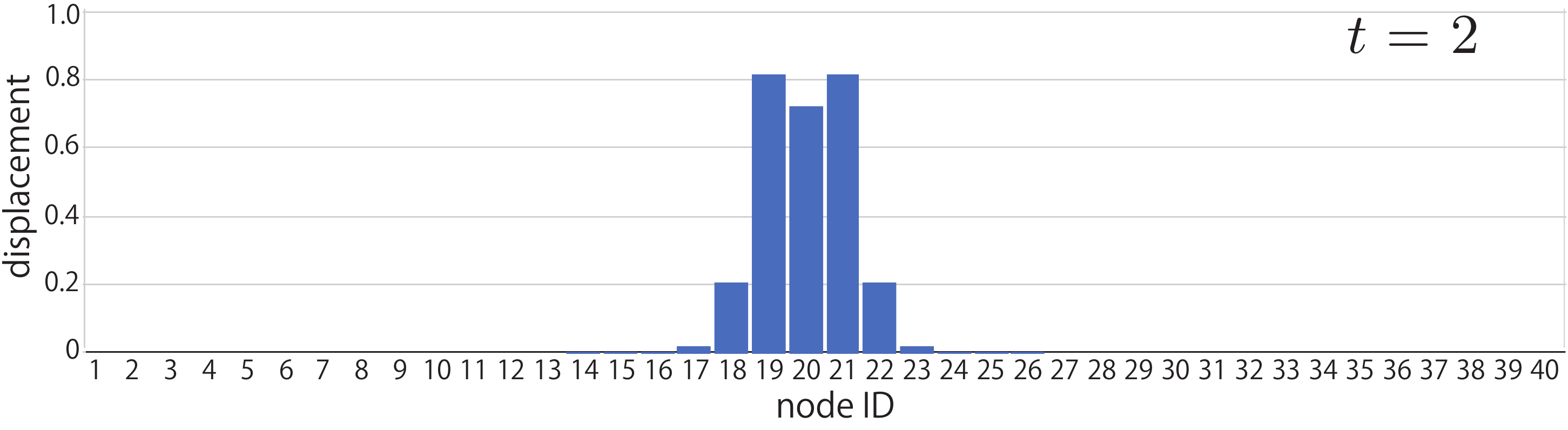}&
\includegraphics[width=0.48\linewidth]{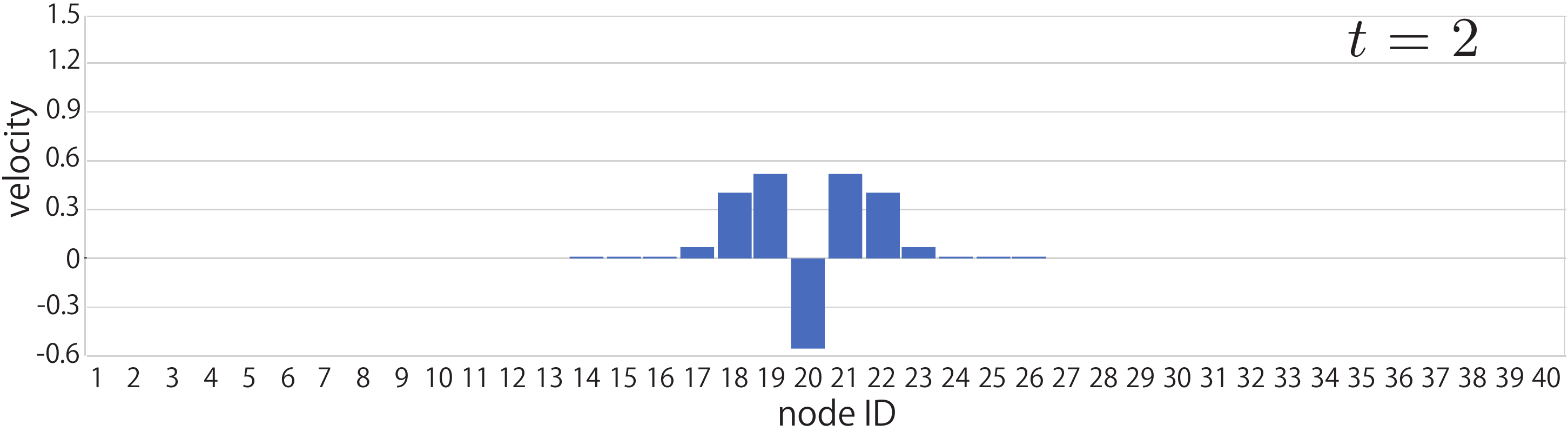}\\
\includegraphics[width=0.48\linewidth]{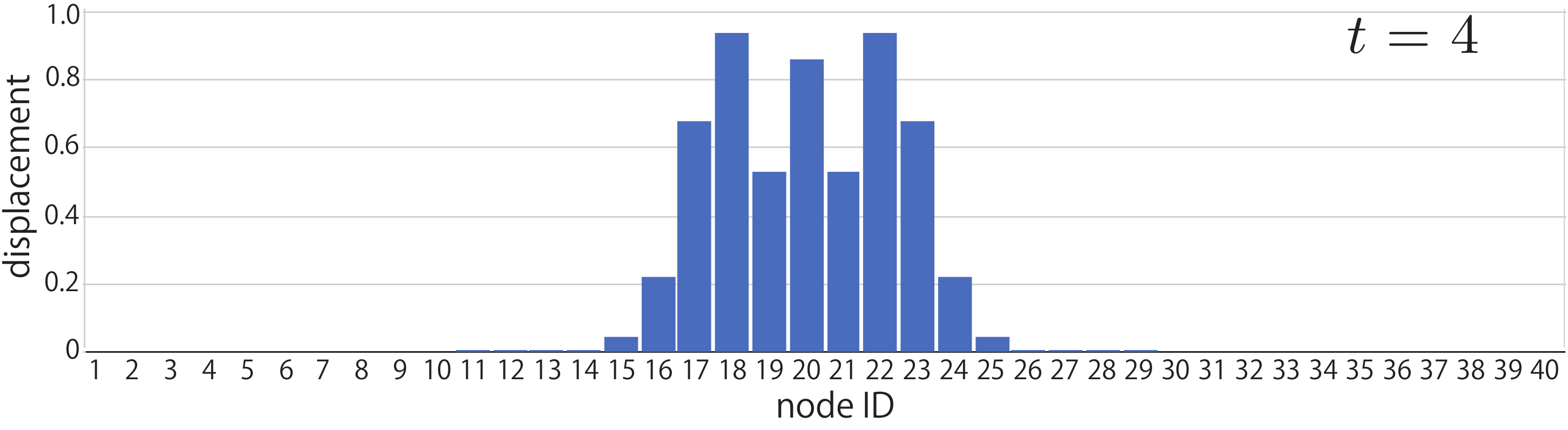}&
\includegraphics[width=0.48\linewidth]{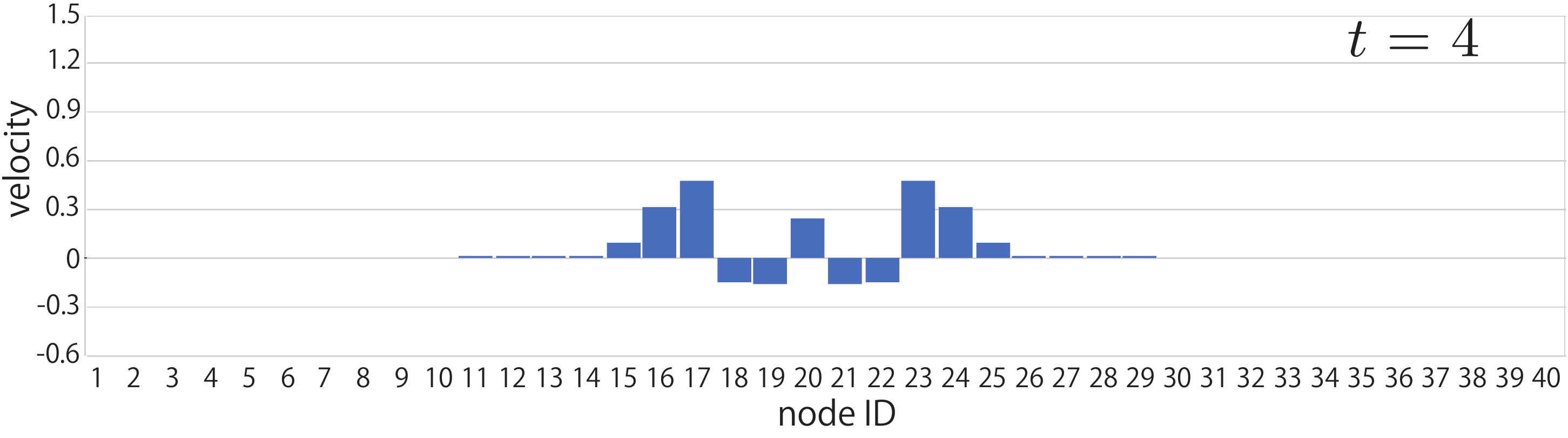}\\
\includegraphics[width=0.48\linewidth]{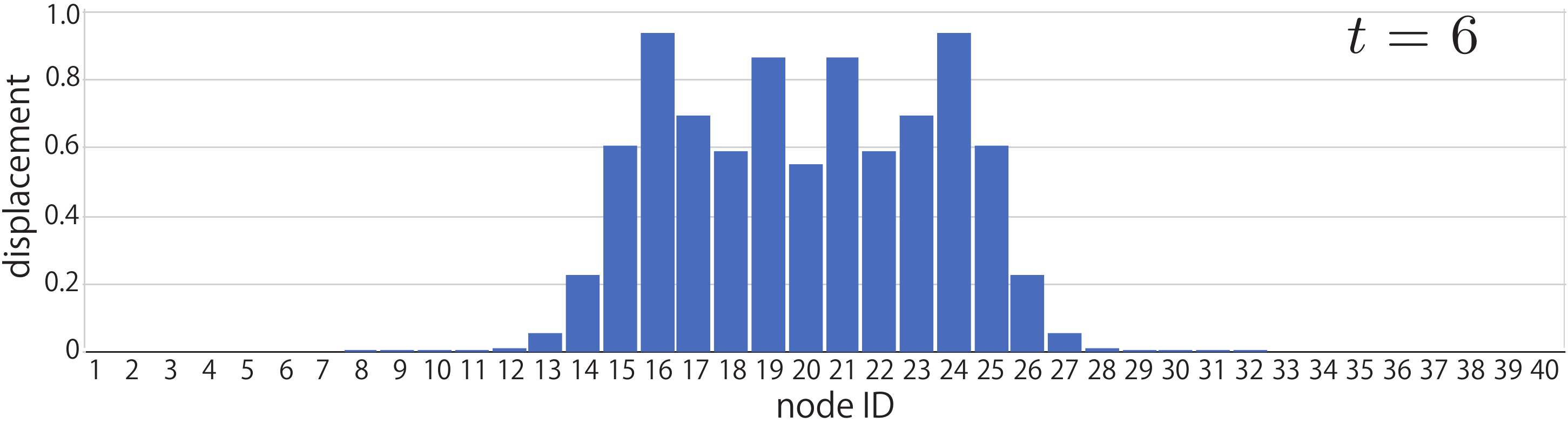}&
\includegraphics[width=0.48\linewidth]{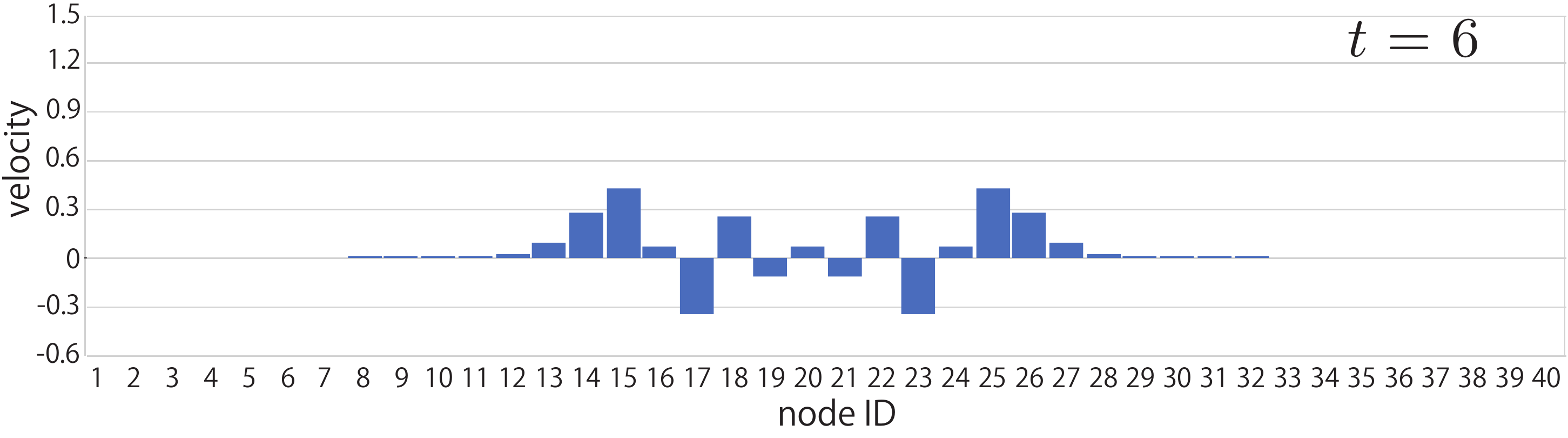}\\
\includegraphics[width=0.48\linewidth]{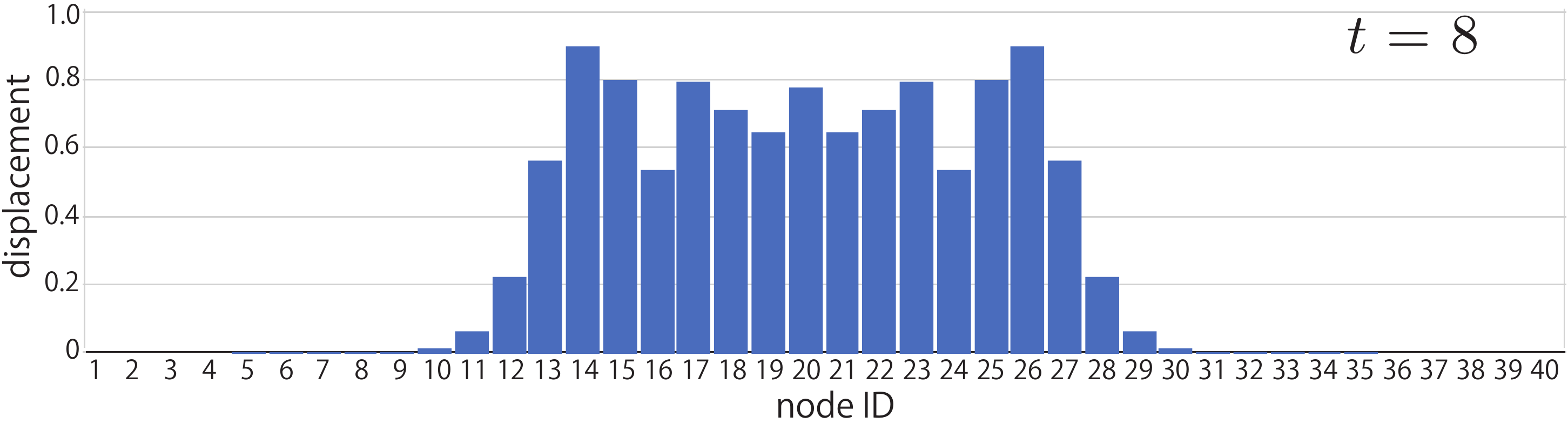}&
\includegraphics[width=0.48\linewidth]{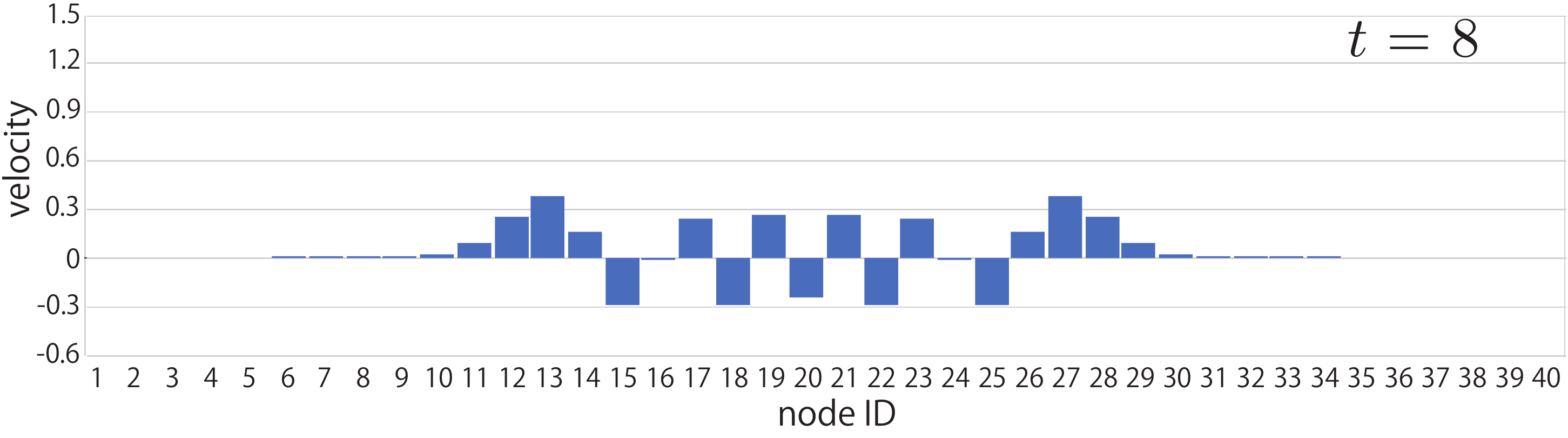}\\
\includegraphics[width=0.48\linewidth]{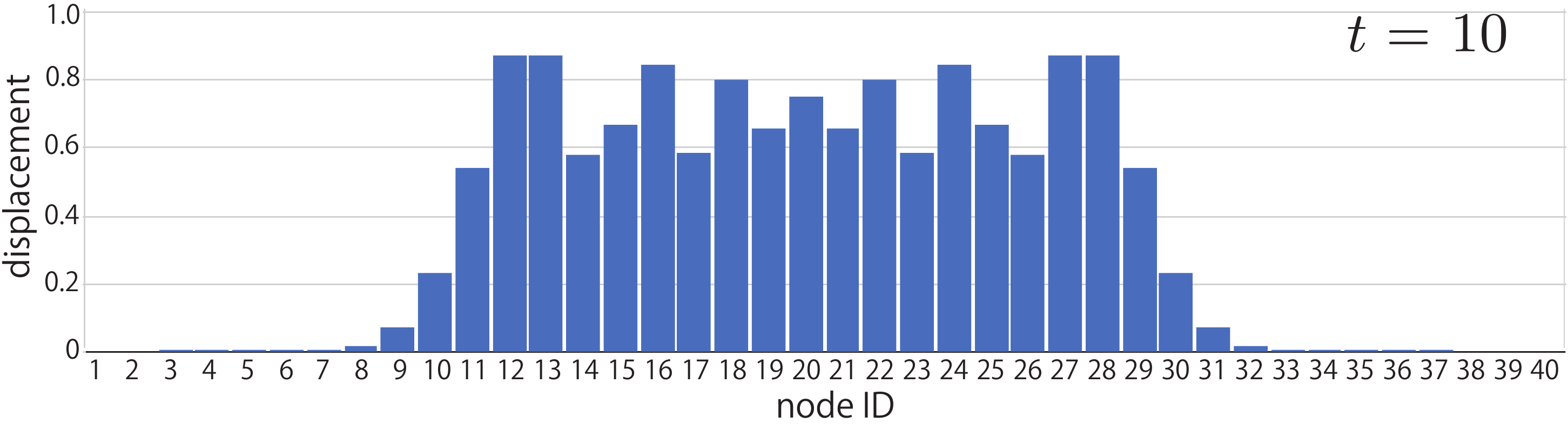}&
\includegraphics[width=0.48\linewidth]{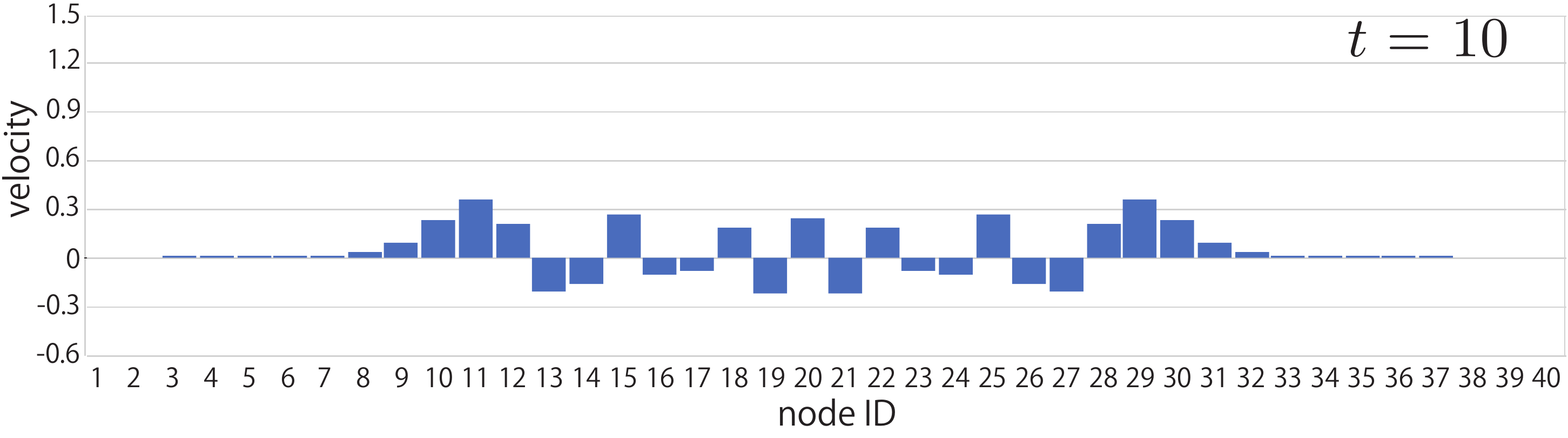}\\
shifted displacement for fermion-type & shifted velocity for fermion-type
\end{tabular}
\caption{Temporal changes of the shifted displacement (left) and the shifted velocity (right) for the fermion-type fundamental equation}
    \label{fig:shifted}
\end{figure}

On the other hand, in the fermion type, one peak can be seen at node $20$ at time $t = 0$, but in fact, the other nodes have negative values close to $0$, and all of these negative values are the same.
In Fig.~\ref{fig:displ}, the reason that the fermion-type displacement occurs in the entire network from the initial stage is that all nodes have initial velocities.
In the first place, the initial velocity is given only to node $ 20 $ on the assumption that nodes other than node $20$ are stationary, but which node is stationary depends on how the coordinate system is taken.
Looking at the results from that standpoint, the velocity distribution that appears here is thought to represent the coordinate system with the origin of the entire center of gravity.
In other words, with the center of gravity of the entire node as the origin, it is considered that the node $20$ has an initial velocity in the positive direction, and at the same time, each of the other nodes has a slight initial velocity in the negative direction.

To confirm this, in Fig.~\ref{fig:shifted}, the fermion type result is set to the coordinate system in which the node $0$ is stationary, considering the initial velocity of the node $0$ and the time change of the displacement obtained from it. 
The left panel shows the time change of displacement of nodes, and the right panel shows the time change of the velocity of nodes.

From the change in displacement of Fig.~\ref{fig:shifted}, it can be seen that the influence propagates at a finite speed from the node $20$.
It can also be seen that only the node $ 20 $ has a value for the initial velocity, and the influence propagates at a finite velocity thereafter.
From the above, it seems that the fermion-type solution is propagating at a remarkable speed at first glance, but the problem can be solved by performing a coordinate transformation from the resting coordinate system of the center of gravity.

It is also noteworthy that the initial velocity that the velocity is given only to the node $20$ is realized as originally expected.
In other words, in the fermion-type, by selecting $\bm{x}^+(0)$ and $\bm{x}^-(0)$, the initial condition of the original wave equation \eqref{eom} can be determined as desired.
The reason why the origin of the coordinates shifts is as follows.
At the initial velocity \eqref{init_veloc_f} of the fermion type, since 
\[
\bm{\mho} \, \bm{\Omega} = \mathrm{diag}(0,\overbrace{1,\,1,\,\dots,\,1}^{\text{$n-1$ components}}),
\]
the information of the eigenvector associated with the eigen angular frequency $\omega_0=0$ in $\bm{x}^+(0)-\bm{x}^-(0)$ is deleted and reflected in the initial condition $\bm{x}_\mathrm{f}(0)$.

From the above consideration, the following can be understood.
The boson-type fundamental equation cannot properly control the initial velocity of the original wave equation even if the initial condition $\bm{x}^+(0)$ and $\bm{x}^-(0)$ are adjusted. 
This means that understanding the causal relationship between user dynamics and the OSN structure and giving the initial conditions of the wave equation appropriately are incompatible in the framework of the boson-type fundamental equation.
On the other hand, in the fermion-type fundamental equation, the initial condition of the original wave equation can be properly adjusted by adjusting the initial condition $\bm{x}^+(0)$ and $\bm{x}^-(0)$.
Therefore, it can be seen that the fermion-type fundamental equation is a better form than the original wave equation and boson-type fundamental equation as a framework for understanding the causal relationship between user dynamics and the OSN structure.

\section{Conclusion}
\label{sec.7}
This paper has derived the closed-form solution of the fermion-type fundamental equation, which is different from the closed-form solution of the boson-type fundamental equation.
In addition, we have derived a closed-form solution of the original wave equation from the solution of the fermion-type fundamental equation. 
The solution of the original wave equation derived from the fermion-type fundamental equation looks in a different form compared with the well-known general solution of the original wave equation directly obtained from the boson-type fundamental equation.

Both solutions derived from two fundamental equations give general solutions of the original wave equation, although the actual solutions for the given initial condition are different.
Specifically, if the initial velocity is $0$ under the initial conditions, both solutions are the same, but otherwise, they are different.
To clarify the difference, we compared the two solutions under the condition that the initial displacement is $0$ and the initial velocity is given only to a specific node.
As a result, it was found that the boson-type solution cannot properly control the initial conditions of the original wave equation by adjusting the initial conditions of the fundamental equation.
Since the fundamental equation is necessary to understand the causal relationship of user dynamics, it means that the control of the initial condition of the original wave equation and the description of the causal relationship is incompatible in the framework of the boson-type fundamental equation.
On the other hand, the fermion-type solution can control the initial conditions appropriately and is compatible with the description of causality.
Therefore, it can be seen that the fermion-type fundamental equation is superior to the original wave equation and boson-type fundamental equation as equations that describe online user dynamics.

\section*{Acknowledgment}
This research was supported by Grant-in-Aid for Scientific Research (B) No.~19H04096 (2019--2021) and No.~20H04179 (2020--2022) from the Japan Society for the Promotion of Science (JSPS), and TMU local 5G research support.




\end{document}